\documentclass[natbib=false,manuscript,screen,nonacm]{acmart}

\usepackage{caption}
\usepackage{subcaption}
\usepackage{algorithm}
\usepackage{algpseudocode}
\usepackage[backend=bibtex,style=authoryear]{biblatex}
\makeatletter
\let\@authorsaddresses\@empty
\makeatother

\AtBeginDocument{%
  \providecommand\BibTeX{{%
    \normalfont B\kern-0.5em{\scshape i\kern-0.25em b}\kern-0.8em\TeX}}}

\begin{document}

\title{Real-Time Scene Reconstruction using Light Field Probes}

\author{Yaru Liu}
\affiliation{%
  \institution{University of Cambridge}
  \country{United Kingdom}
}

\author{Derek Nowrouzezahrai}
\affiliation{%
  \institution{McGill University}
  \country{Canada}
}

\author{Morgan McGuire}
\affiliation{%
  \institution{Roblox and McGill University}
  \country{Canada}
}





\begin{abstract}
Reconstructing photo-realistic large-scale scenes from images, for example at city scale, is a long-standing problem in computer graphics. Neural rendering is an emerging technique that enables photo-realistic image synthesis from previously unobserved viewpoints; however, state-of-the-art neural rendering methods have difficulty efficiently rendering a high complex large-scale scene because these methods typically trade scene size, fidelity, and rendering speed for quality. 
The other stream of techniques utilizes scene geometries for reconstruction. But the cost of building and maintaining a large set of geometry data increases as scene size grows.
Our work explores novel view synthesis methods that efficiently reconstruct complex scenes without explicit use of scene geometries. 
Specifically, given sparse images of the scene (captured from the real world), we reconstruct intermediate, multi-scale, implicit representations of scene geometries. 
In this way, our method avoids explicitly relying on scene geometry, significantly reducing the computational cost of maintaining large 3D data.
Unlike current methods, we reconstruct the scene using a probe data structure. Probe data hold highly accurate depth information of dense data points, enabling the reconstruction of highly complex scenes. By reconstructing the scene using probe data, the rendering cost is independent of the complexity of the scene. 
As such, our approach combines geometry reconstruction and novel view synthesis. 
Moreover, when rendering large-scale scenes, compressing and streaming probe data is more efficient than using explicit scene geometry. Therefore, our neural representation approach can potentially be applied to virtual reality (VR) and augmented reality (AR) applications.
\end{abstract}

\begin{teaserfigure}
\centering
\includegraphics[width=15cm]{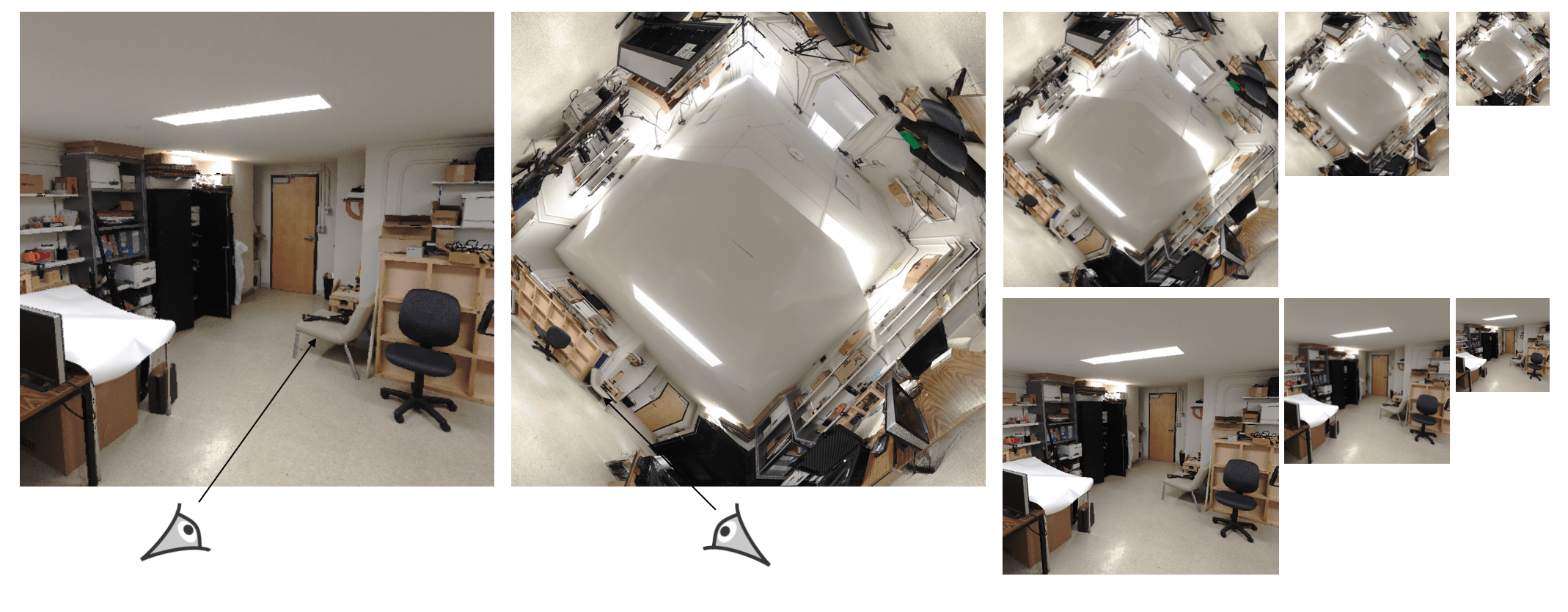}
\caption[]{A 3D scene and its probe-based parameterizations. Left to right: ray tracing the scene via scene geometries; ray tracing the scene using our probe data; hierarchical scene representation.}
\label{fig:2.1}
\end{teaserfigure}


\maketitle
\section{Introduction}
In computer graphics, the photorealistic synthesis of three-dimensional (3D) scenes has long been a focal point of research and development. Approaches to scene generation involve using surface representations, such as point clouds, meshes, and surfaces, and volume representations, such as voxels and density, to represent objects and their environments [\cite{aliev_neural_2020}, \cite{burov_dynamic_2021}, \cite{carr_reconstruction_2001}, \cite{sitzmann_deepvoxels_2019}, \cite{genova_local_2020}]. While these methods have achieved remarkable progress in capturing visual realism, they often suffer from limitations such as computational complexity and difficulty handling complex deformations and intricate geometry. Moreover, these representations suffer from high memory requirements and have difficulty scaling to large scenes, for example, a city-scale scene, because creating and maintaining a large 3D data set is computationally expensive.

To overcome these limitations, many approaches have been taken to represent and render 3D scenes without the need for explicit mesh structures [\cite{nerf}, \cite{fastnerf}, \cite{BungeeNeRF}, \cite{wildnerf}], offering unique advantages and paving the way for high-performance graphics applications.


One approach is to reconstruct scenes using two-dimensional (2D) representations, for example, images [\cite{nerf}, \cite{BungeeNeRF}, \cite{wildnerf}]. Neural rendering combines classical computer graphics and machine learning to synthesize images from real-world observations and has been gaining more popularity in recent years. Many works in this field focus on 3D scene reconstructions using 2D representations and have made significant progress in conquering limitations of scene size, rendering speed, and fidelity [\cite{PlenOctrees}, \cite{fastnerf}]. However, it is challenging to render complex, large-scale scenes efficiently.

We will explore synthesizing a complex room-scaled scene without using explicit scene geometry. 
What is novel about our method is that we generate high-quality probe data [\cite{gi-probes}] using real-world data and have created a mix of 2D and 3D scene representation. We will demonstrate that representing a scene using probe data is highly memory efficient. Moreover, we propose simulated probe data to enable high-quality rendering in real-time when traversing the scene.
\section{Related Work}

\textbf{Light Field Probes}
A light field is the total amount of light in three-dimensional space at any position and in any direction
[\cite{gi-probes}] encodes additional information about the scene geometry into light field probes and performs accurate world-space ray tracing for glossy and near-specular indirect transport. In addition, [\cite{gi-probes}] extends prefiltered (ir)radiance maps with visibility-aware sampling and interpolation to eliminate light and dark leaks.
Many shading algorithms leverage the efficient query and sampling operations exposed by the light field probe representation to synthesize high-fidelity outputs at high-performance rates. For example, [\cite{majercik_dynamic_2019}] inherit the advantages of the probe representation used in [\cite{gi-probes}] and extend to treat dynamic geometry and lighting variations at runtime. 
[\cite{majercik_scaling_2021}] refines [\cite{majercik_dynamic_2019}] to accelerate the evaluation of global illumination. [\cite{majercik_scaling_2021}] replaces an expensive computation of diffuse global illumination with the light field probe, which only takes $O(1)$ to look up into the data structure. In addition, [\cite{majercik_scaling_2021}] reuses probes to sample incident radiance for glossy global illumination and combines with filtered screen-space and geometric glossy ray tracing.
Therefore, light field probes can be used in the context of both radiance lookups and world-space ray tracing.

\bigskip
\noindent \textbf{Large Scale Scene Reconstruction}
Conventional large-scale 3D reconstruction rely on many mature methods, for example structure from motion, and most of these methods jointly optimize a set of 3D points and camera poses to maintain consistency with image features, e.g. SIFT (Figure 2a), extracted from a large collection of input images [~\cite{agarwal_building_2009}]. Motivated by Neural Radiance Fields (NeRF) [~\cite{nerf}] (Figure 2c), a series of follow-up works have proposed techniques for rendering large scenes, spanning from street view~\cite{rematas_urban_2021} to building- and city-scale, and even earth-scale [~\cite{BungeeNeRF}, ~\cite{tancik_block-nerf_2022}]. These methods usually partition a large scene into smaller pieces and use localized NeRFs in each spatial cell [~\cite{tancik_block-nerf_2022} ~\cite{turki_mega-nerf_2022}], or learn multi-scale representations of a 3D scene [~\cite{BungeeNeRF}. To render such large-scale scenarios, changing illumination and transient occluders usually need to be carefully handled due to the unstructured input data. "NeRF in the Wild" [~\cite{wildnerf}] learns appearance embeddings per training image to handle the inconsistent scene appearance. In contrast, representing scenes with probe data and with multiple hierarchies of fidelity, our proposed method will be able to synthesize high-fidelity details of large-scale scenes with complex geometries and at various resolutions.      

\bigskip
\noindent \textbf{Fast Rendering}
NeRFs can render scene content from unobserved viewpoints, but they do not scale to large scenes. Firstly, they require prohibitively large computation during rendering – synthesizing a single pixel requires sampling the network hundreds of times. Secondly, NeRFs suffer from long inference time because changing the viewing directions leads to resampling different points along each pixel’s view ray. Training a NeRF can also be time-consuming due to the expensive ray-casting and optimization process, taking up to days to model a scene on a single GPU. Many existing works approach these limitations by, e.g., storing precomputed features in a separate data structure [~\cite{PlenOctrees}
~\cite{hedman_baking_2021}, ~\cite{garbin_fastnerf_2021}, ~\cite{neuralode}], like a sparse voxel tree, leveraging spatial partitioning as in "MegaNeRF" [~\cite{turki_mega-nerf_2022}], "KiloNeRF" [~\cite{reiser_kilonerf_2021}], and "DeRF" [~\cite{rebain_derf_2020}], and tailoring the sampling to more efficient strategies [~\cite{neff_donerf_2021}].


\section{Overview}

\begin{figure}[H]
     \centering
     \begin{subfigure}[b]{0.5\textwidth}
         \centering
        \includegraphics[width=\textwidth]{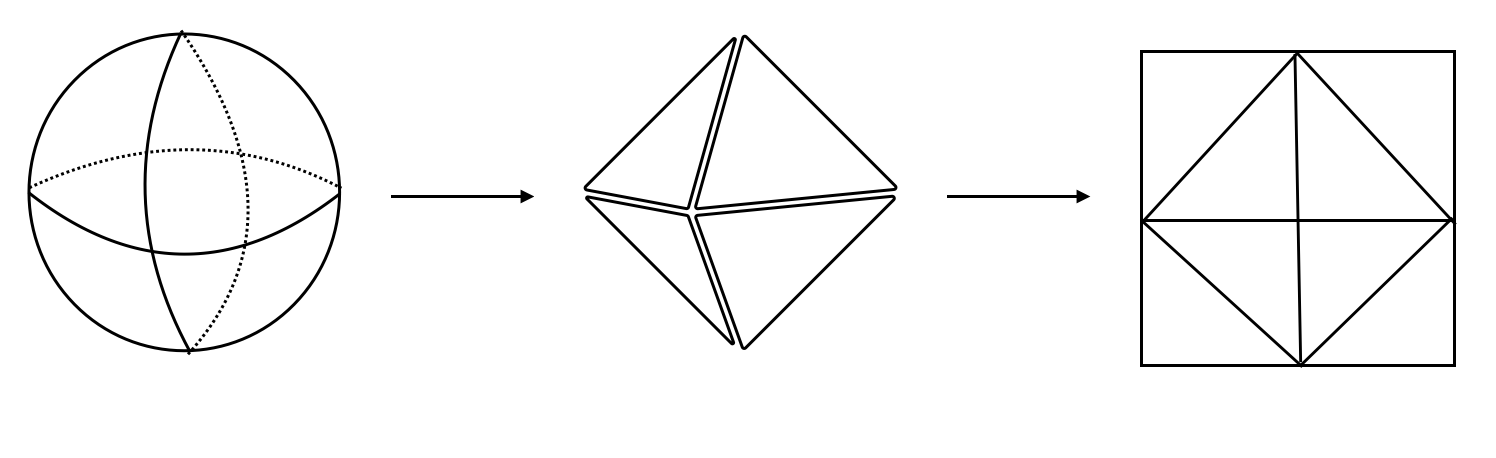}
        \subcaption{}\label{fig:2.1}
    \end{subfigure}
    \begin{subfigure}[b]{0.3\textwidth}
         \centering
        \includegraphics[width=\textwidth]{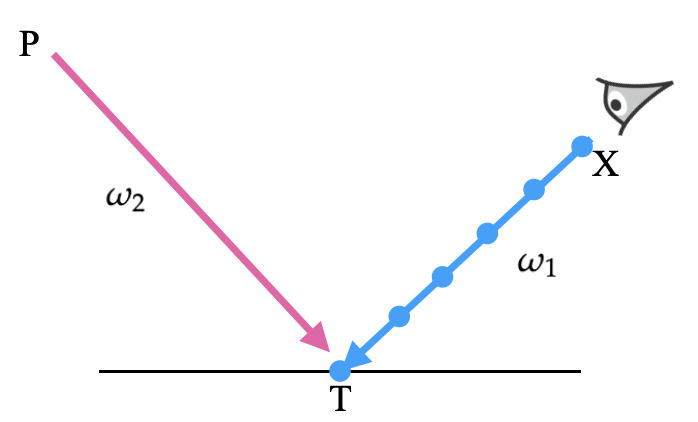}
        \subcaption{}\label{fig:2.0}
    \end{subfigure}
        \caption[]{(a) A visualization of the octahedral mapping; (b) A visualization of Ray marching.}
        \label{fig:3.0}
\end{figure}

We encode all scene information into light field probe data. Using octahedral mapping (Figure ~\ref{fig:2.1}), the 3D environment is first projected onto a unit sphere. 
Then, everything on the unit sphere is projected onto an octahedron. Next, octants are unfolded to map the octahedron to a 2D texture image that is later used during ray tracing. Octahedral mapping reduces distortion at the poles and requires less memory and storage compared to cubemaps [\cite{Blinn1TextureAR}] and therefore is ideal for representing light probes. 

The probe data is used for ray queries. 
Let $O$ be a probe origin and $(X, \omega_1)$ be a ray where $X$ is the ray origin in the 3D scene and $\omega_1$ is the direction that the camera looks at in the scene (Figure ~\ref{fig:2.0}). Assume $X \neq P$. To find out the irradiance values $L_o(X, \omega_1)$, we want first to find out the intersection of the ray $(X, \omega_1)$ with the scene surfaces. 
The intersection $T$ (Figure ~\ref{fig:2.0}) is the nearest point on surfaces that can be reached by stepping from $X$ along the direction $\omega_1$. We will explain the ray-scene intersection algorithm in Section ~\ref{4.2}. Let $\omega_2 = \overrightarrow{PT}$. Once $T$ is found, we check if the probe can see $T$ by checking if $r_p(\omega_2) = d(P, T)$ holds. If the probe has visibility of T, the irradiance value at point $X$ can be obtained as 
\mbox{$L_o(X, \omega_1) = L_o(P, \omega_2) = L_p(\omega_2)$}.

To perform ray queries on a probe, we first project a ray onto an octahedral map. Then we conduct ray marching along the projected ray. Each probe captures distance and irradiance information. The distance information encoded in the octahedral map enables us to find the ray-scene intersection. Once the intersection is found, we can look up the octahedral map to query irradiance values and write the color to the image buffer. We can use Musgrave's height field tracing algorithm [\cite{musgrave}] to enhance the performance of ray marching. We use a chain of octahedral maps of different resolutions to resolve the performance issue. During tracing, we first use lower-resolution probe data to skip over empty space along the ray because a single pixel covers a box of pixels in the high-resolution octahedral maps. Once we find a potential intersection, we use higher-resolution maps for finer tracing.  
\begin{figure} [H]
     \centering
     \begin{subfigure}[b]{0.26\textwidth}
         \centering
        \includegraphics[width=\textwidth]{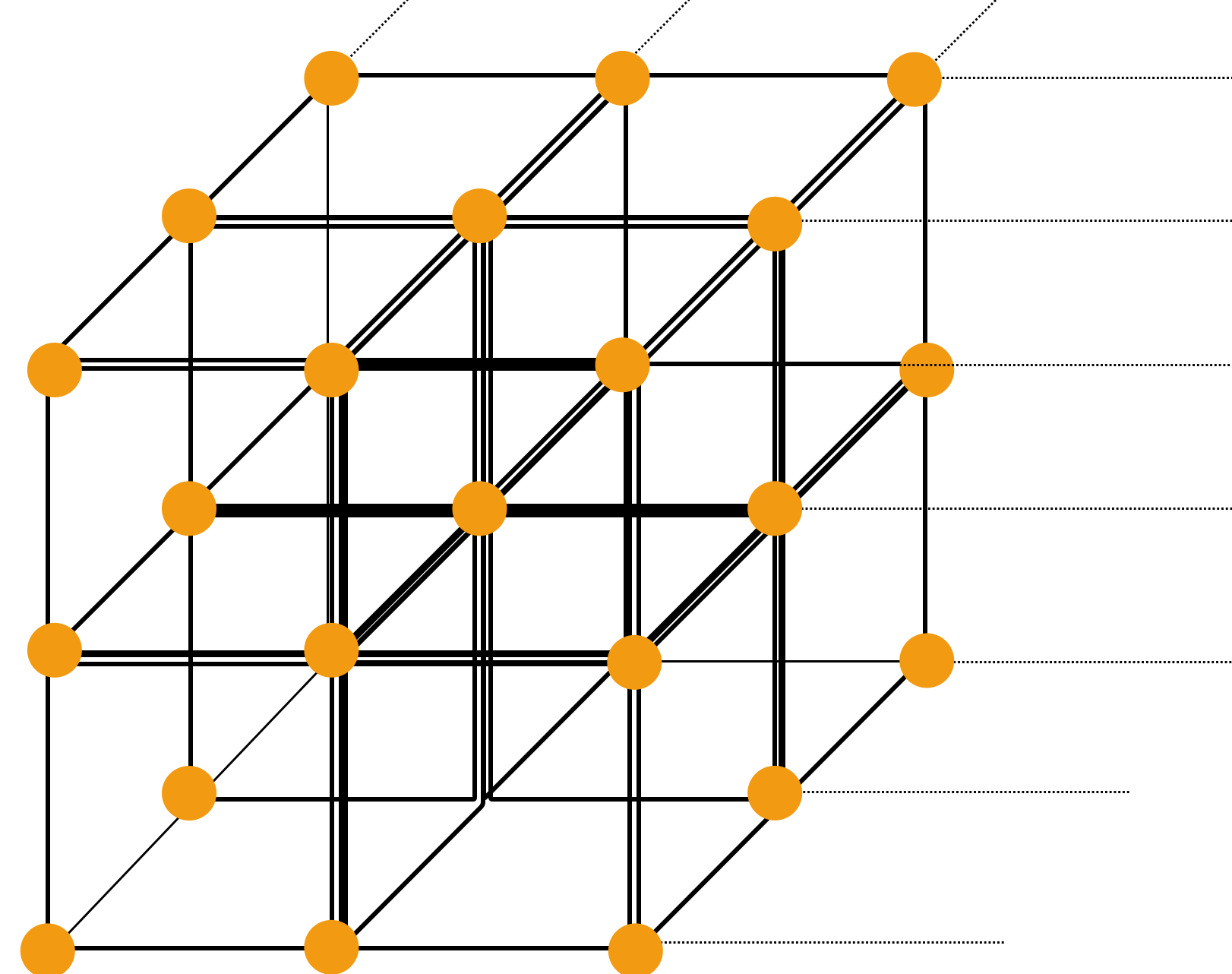}
        \subcaption{}\label{fig:4.2b}
    \end{subfigure}
    \begin{subfigure}[b]{0.31\textwidth}
         \centering
        \includegraphics[width=\textwidth]{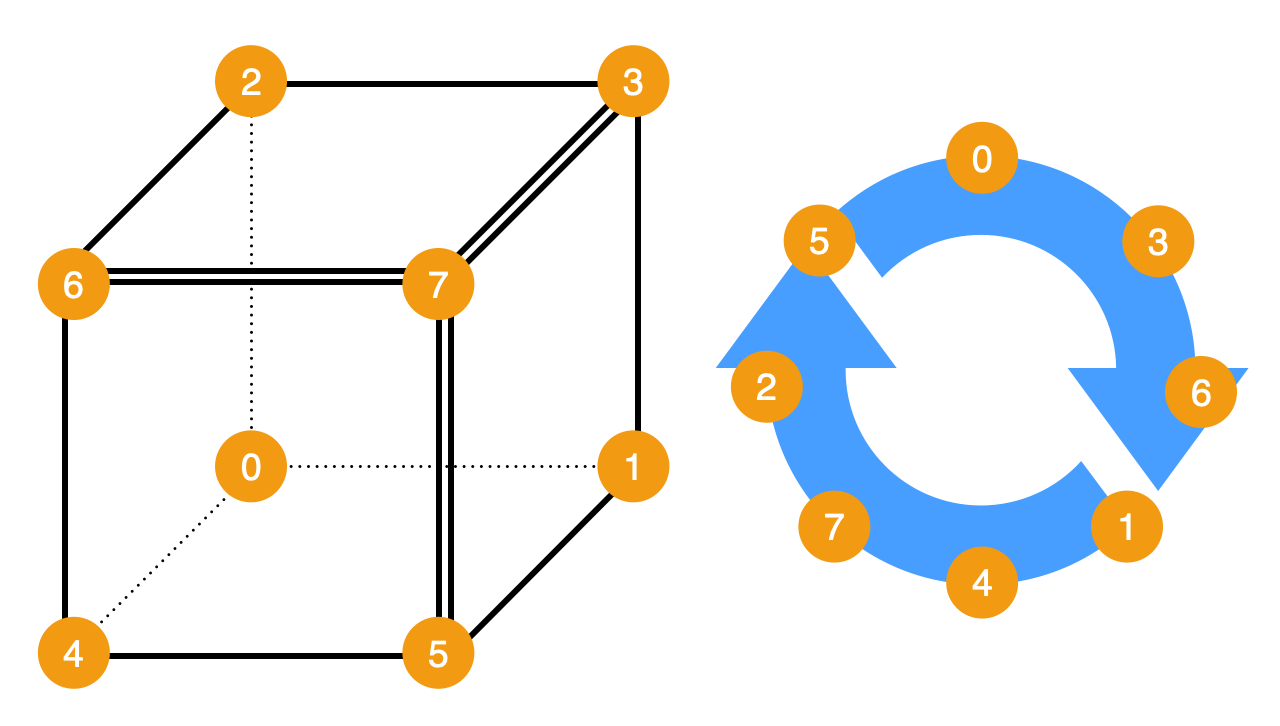}
        \subcaption{}\label{fig:2.6}
    \end{subfigure}
    \begin{subfigure}[b]{0.31\textwidth}
         \centering
        \includegraphics[width=\textwidth]{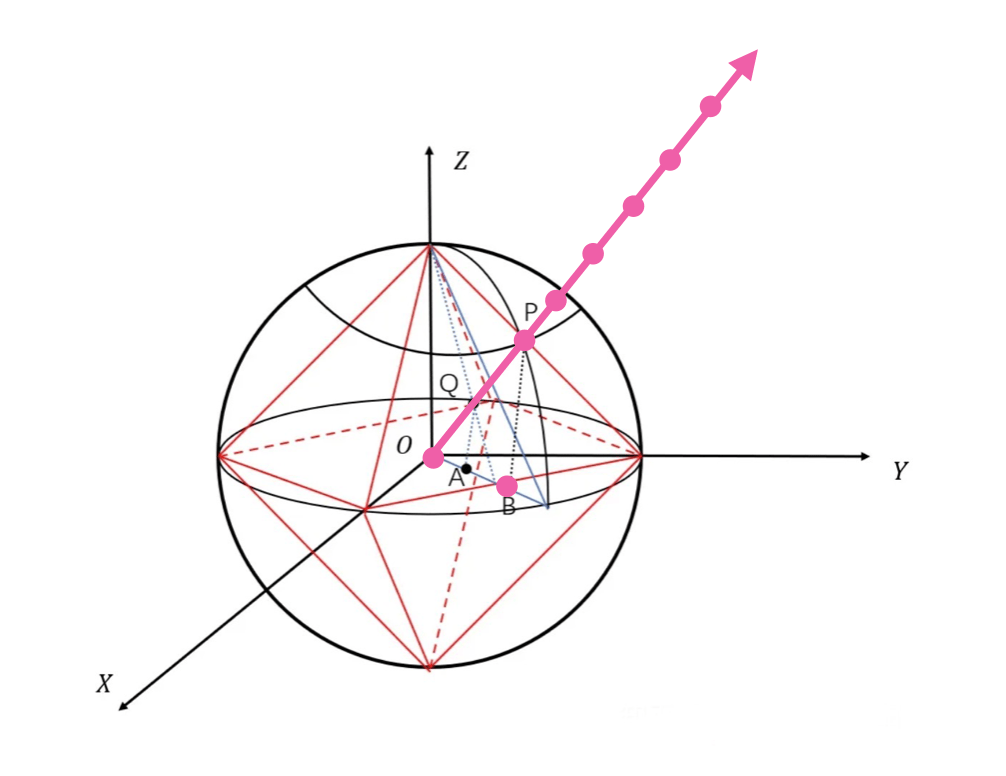}
        \subcaption{}\label{fig:4.10}
    \end{subfigure}
        \caption[]{
        (a) probes are placed vertices of a uniform 3D grid, which consists of cubes; (b) an indexed cube with probes at vertices. Arrows show the order in
which probes are traced in 8-probe scenarios; (c) A ray is shoot from probe origin $O$. All points along the ray are mapped to point $P$ on the unit sphere and finally projected to point $B$ on the octahedral map.}
        \label{}
\end{figure}


\label{multiple-probes}
When the scene size or complexity of 3D scenes grows, more than one probe is required to reconstruct the entire scene. We place probes at the vertices of a uniform 3D grid (Figure~\ref{fig:4.2b}). Our goal is to find the cube (Figure~\ref{fig:2.6}) that encloses the ray-scene intersection point. Given the cube, we follow the previous work [\cite{gi-probes}] to trace irradiance value by looping over 8 probes. 
The problem of backfaces is common in rendering. 
Given a surface point and an eye ray, if the dot product of its surface normal and a viewing direction is above $0$, the surface point is invisible to the eye because it points away from the ray origin. Therefore, surface normals are required for checking visibility for ray queries on a probe. We encode the normal information into the probe data so our probe representation is robust to light leak.


\section{Methodology}
\label{section5}


\subsection{Simulated Probe Data}

When the eye's position is further away from the position of the probe that it traces against, the reconstruction quality becomes worse. To resolve the issue, we can use more probes to ensure a probe is close enough to the eye position. However, additional probe data largely increase memory usage and the computation cost of data processing. So we propose a novel simulation method to enhance reconstruction quality without collecting unnecessarily excessive data.
The probe data is generated using real-world data captured by a FARO scanner. We first place the scanner at a few locations to capture the scene that we want to reconstruct.   Next, during data processing, we create a project point cloud which consists of all scans, and unify the coordinate systems of different scans. In this way, we can pick any point in the location as the probe location. Then, we map all points to the unit sphere of the probe, project to the octahedron and finally create an octahedral map. In this way, we are able to generate as many probes as needed for reconstructing a high-quality scene.

\subsection{Ray Tracing Using One Probe}
\label{4.2}
\noindent \textbf{Ray Tracing}
The ray tracing algorithm is built upon the previous work [\cite{gi-probes}]. We first project a ray onto an octahedral depth map and obtain at most 4 ray segments. We then march along each ray segment step by step to find where the ray intersects with the scene and check visibility using normal maps [Algorithm \ref{fig:alg2}]. If the intersection point is visible to the probe, we look up the irradiance value in the irradiance map and write to the image buffer. 

To apply hierarchical tracing, we first use lower-resolution maps to skip over empty space along the ray [Algorithm \ref{fig:alg3}]. We call the process low-resolution tracing and compare distance values to find potential intersection points. 
It is worth noting that the low-resolution tracing only checks for intersections and does not check whether the intersection point is on a back face or visible to the probe. If the low-resolution tracing fails to find the intersection, we call the tracing result MISS and continue to trace the next ray segment. Otherwise, we move forward to trace using the higher-resolution map [Algorithm \ref{fig:alg4}]. 
The high-resolution tracing is similar to the low-resolution tracing but different in that we need to check visibility using the normal map to ensure that the intersection does not lie on a back surface. The probe does not know whether the ray segment intersects the scene at this point because the ray can shoot from below the surface. We call this tracing result UNKNOWN. If the current intersection passes the visibility test, we call the tracing result HIT and find its corresponding irradiance value.

\bigskip
\noindent \textbf{Eye Aligned with a Probe}
When the eye aligns with a probe during reconstruction, we propose a novel ray tracing algorithm that only takes $O(1)$ cost and avoids hierarchical tracing. We find that when a probe shoots a ray into the 3D scene, all points along the way are mapped to the same point on the unit sphere so all points along the ray are projected to the same point on the octahedral (Figure~\ref{fig:4.10}). Therefore, rather than ray marching, we can directly project the normalized ray onto the octahedral map and look up its irradiance value [Algorithm\ref{fig:alg5}]. In this way, we can omit the expensive ray-scene intersection test and largely improve the algorithm's efficiency. 
Our work is implemented in Python and PyOpenGL, and run on NVIDIA GeForce RTX 4080 GPU.

\subsection{Ray Trace Using Multiple Probes}
\label{6.4}

\textbf{Less Than 8 Probes}
If a scenario has more than 1 but less than 8 probes, such as a 2-probe or 4-probe scenario, we start from the probe that is closest to the eye. If the trace result is MISS or UNKNOWN, rather than just UNKNOWN, we loop to the next closest probe until all probes are used up. 
Our experiments show that usually, if the first two probes closest to the ray origin cannot capture the intersection, neither do the rest of the probes. 

\bigskip
\noindent \textbf{Grid Iterator}
When 8 or more probes are required to reconstruct a scene, we place probes at the vertices of a uniform 3D grid (Figure~\ref{fig:4.2b}), which consists of cubes (Figure~\ref{fig:2.6}). Unlike previous work [~\cite{gi-probes}, ~\cite{majercik_dynamic_2019}, ~\cite{majercik_scaling_2021}], our reconstruction does not require scene geometries so it is not trivial to find the ray-scene intersection points. To resolve this problem, we use a grid iterator. The grid iterator iterates over cubes that intersect with the ray until it finds the cube whose probe(s) is able to capture ray-scene intersection points. To be specific, the iterator starts from the first cube that intersects with the ray and then marches along the ray until the ray exits the grid.


\section{Implementation Details}

\subsection{Probe Data Generation}
\label{4.1}
We creatively generate probe data using point cloud data captured from the real world.
To reconstruct a scene using multiple probes, we scan at different positions and then process all scans using FARO's SCENE software. 
During data processing, we first colorize and register all scans and then apply color balancing to ensure color consistency before exporting (Figure~\ref{fig:3.8}). 
Finally, all data are exported to a single file in $xyz$ format. The exported file includes $x, y, z$ coordinates of each point in the world space and its $r, g, b$ value.

With the processed data, we are ready to generate octahedral maps. 
In our experiments, we use two-level hierarchical tracing to enhance ray tracing performance. Thus we need to generate 5 octahedral maps for each probe. 
For a simple scene, octahedral maps with 1024x1024 and 64x64 resolution are able to produce high-quality synthesis. However, higher-resolution octahedral maps produce better results for a more complex scene, and we use 2048x2048 and 128x128 resolutions maps to reconstruct a room-scale scene.

To create octahedral maps, all points are converted from world space to the coordinate system of the probe and then projected to the 2D octahedral map. Usually, more than one point will be projected onto a single image pixel, but we only keep the point that has the smallest Euclidean distance to the probe. 
In this way, we obtain a distance and irradiance map. To encode the normals into octahedral maps, we normalize the vector from the probe origin pointing towards the point cloud. 
It is worth noting that normal is a vector that is perpendicular to the surface, so we do not store typical normals but, more precisely, incoming directions to the probe. However, we call such a map a normal map for simplicity. Low-resolution maps are derived from high-resolution maps by convolution. 

\begin{figure}
     \centering
     \begin{subfigure}[b]{0.23\textwidth}
         \centering
         \includegraphics[width=\textwidth]{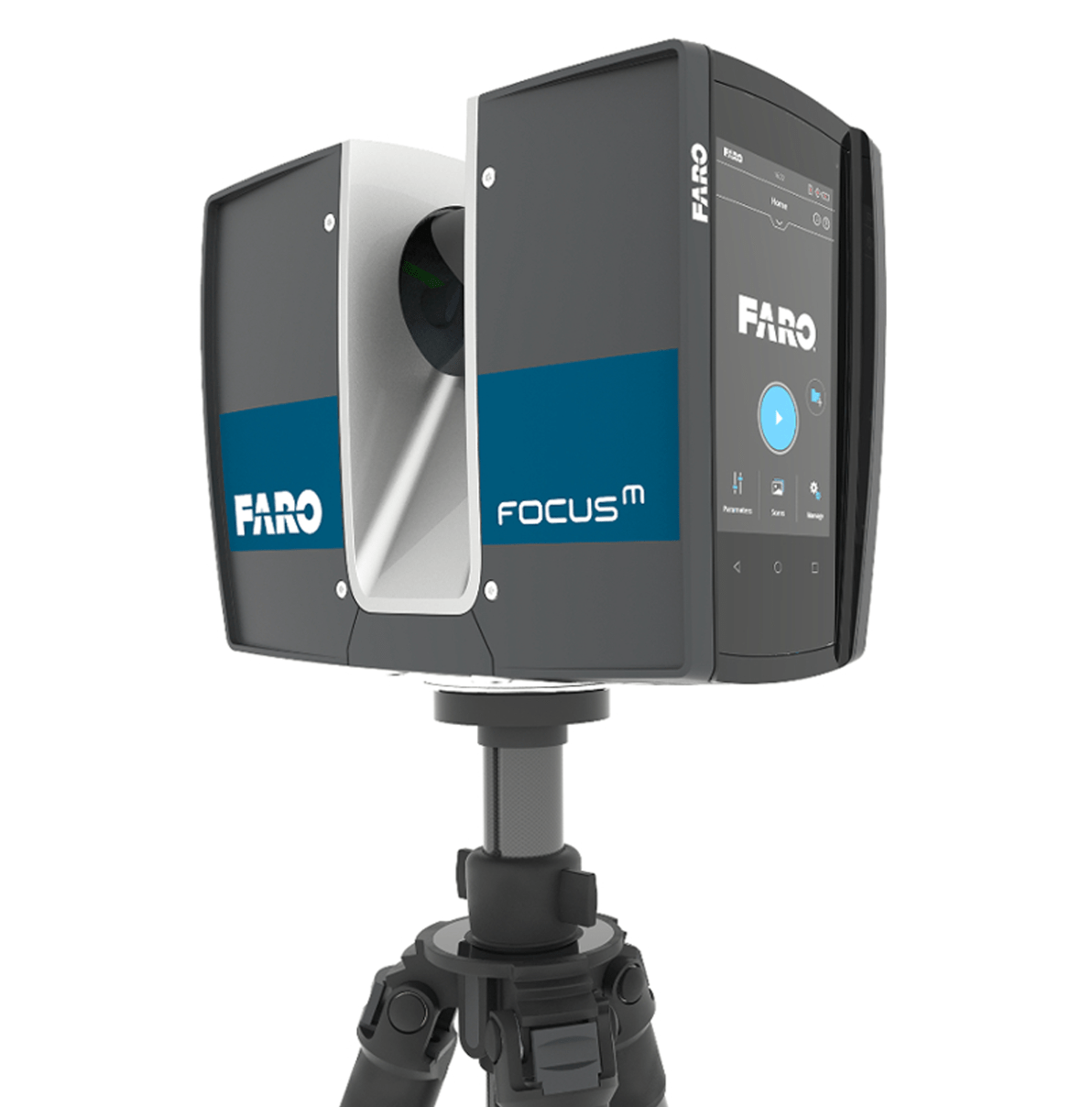}
         \caption{}
         \label{fig:3.1}
     \end{subfigure}
     \begin{subfigure}[b]{0.2\textwidth}
         \centering
         \includegraphics[width=\textwidth]{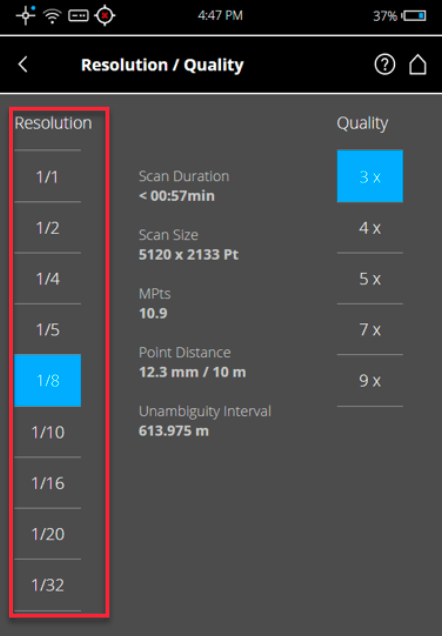}
         \caption{}
         \label{fig:3.3}
     \end{subfigure}
      \begin{subfigure}[b]{0.5\textwidth}
         \centering
         \includegraphics[width=\textwidth]{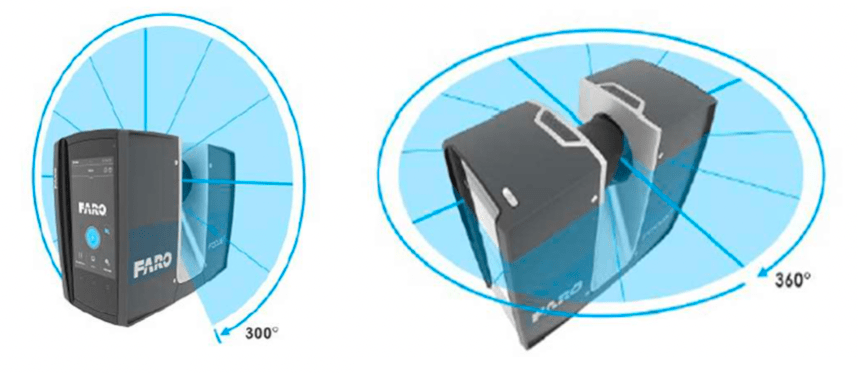}
         \caption{}
         \label{fig:3.2}
     \end{subfigure}
       \caption[Distance and normal octahedral maps]{(a) A FARO scanner; (b) FARO parameter setting UI; (c) the scanner scans an environment; vertical angle covers $360^{\circ}$ and horizontal angle covers $300^{\circ}$. All images $\copyright$ FARO\textsuperscript{\textregistered}.}
        \label{fig:3.13}
\end{figure}
\footnote{All images $\copyright$ FARO\textsuperscript{\textregistered} are from FARO's website and official training materials.} 

\subsection{Data Collection}
 To collect data from the real world, the scanner emits laser beams that bounce off surfaces and return to the scanner. The vertical angle covers $300^{\circ}$ and horizontal rotation covers $360^{\circ}$ (Figure~\ref{fig:3.2}). By scanning the environment in a $360^{\circ}$ sweep, we obtain a set of unstructured dense point cloud data which contains irradiance and depth information, yielding our probe data after proper data processing. The scanner can capture up to 266-megapixel color information and distance with an accuracy of 2 mm at 10m. This accuracy is crucial to generate high-quality octahedral texture maps. 

The parameter setting, which includes resolution and quality, is essential for the data collection (Figure~\ref{fig:3.3}) because the quality of the data determines the quality of probe data, which is critical for rendering high-quality outputs. 
The resolution determines the level of detail during the scanning process and more data is captured under the high-resolution setting. 
The quality affects the level of noise reduction and the scanning time.
The higher the quality setting, the longer the time it takes to scan. Depending on the chosen resolution and quality, a single scanning can take from a minute up to around an hour. 

\subsection{Data Processing} 
The captured point cloud data is loaded into FARO SCENE software for data processing
The scanner first captures the point cloud data in grayscale. To colorize the data, the scanner takes pictures of the environment and applies the colorful pictures to the data.
When reconstructing a scene using multiple probes, we place the scanner at different locations to thoroughly capture the environment. Each scan has its own coordinate system. Therefore, registration is required to combine different scans into a unified coordinate system. Moreover, scans from different viewpoints may have overlapping regions or different levels of noise and errors. By aligning and registering these scans, we can minimize inconsistencies, correct errors, and enhance the overall accuracy of the combined point cloud.
In addition, data captured at different locations have inconsistent color due to variations in lighting conditions and distance to the light sources. 
So we apply color balancing to correct these inconsistencies and biases, making the colors appear more natural and consistent. In this way, we achieve consistent color representations across the entire point cloud, accurately representing colors during rendering, resulting in more realistic and consistent representations.


\section{Results}
\label{section7results}
We present our experimental results of reconstructing a complex lab room (Figure~\ref{fig:5.0}). 
The lab room has a rectangular shape with a width of around 3 meters and a length of around 6 meters. The results show that our methodology successfully synthesizes a highly complex room-scaled scene. In addition, the creative mixture of 2D and 3D representation is able to efficiently reconstruct a 3D scene. Moreover, the reconstruction time is independent of the complexity of a scene.


\subsection{Probe Data}
\label{5.1}
\textbf{Visualization of Probe Data}
We generate probes using real-world data captured by FARO scanner. We set the scanner parameters to resolution $\frac{1}{5}$ and quality $4x$ and conduct 4 scans to capture the scene information. Each scan takes around 7 min and yields approximately 28 million point cloud data for a single scan. Given that we use 2-level hierarchical tracing, each probe generates 5 octahedral maps (Figure~\ref{fig:5.0}).

\begin{figure}[th]
\centering
\includegraphics[width=14cm]{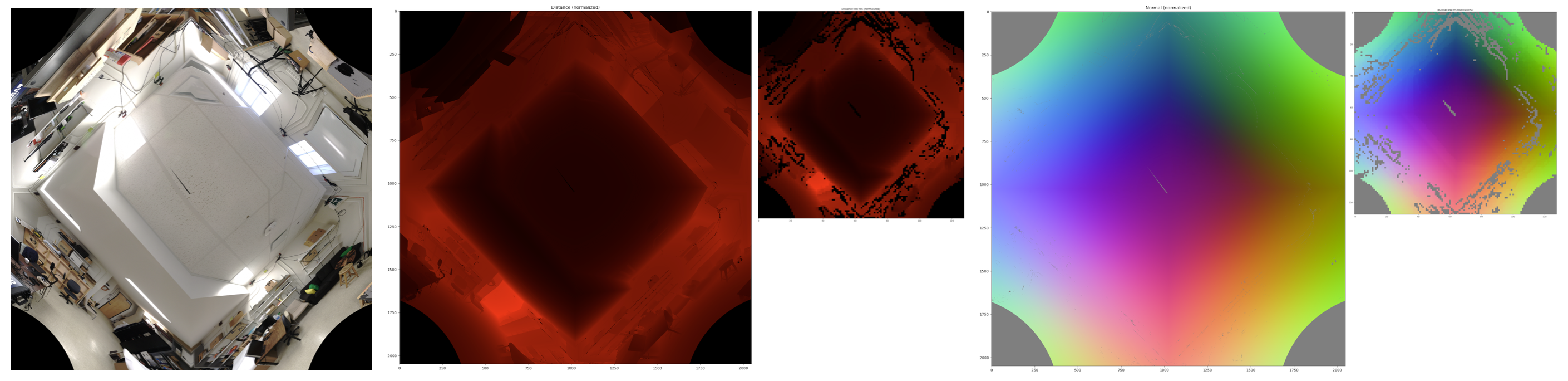}
\caption[5 octahedral maps generated for a single probe]{An example of 5 octahedral maps generated for a single probe. Left to right: irradiance map of 2048x2048, distance maps of 2048x2048 and 128x128, normal maps of 2048x2048 and 128x128.}
\label{fig:4.1}
\end{figure}

\begin{figure}
     \centering
     \begin{subfigure}[b]{0.29\textwidth}
         \centering
         \includegraphics[width=\textwidth]{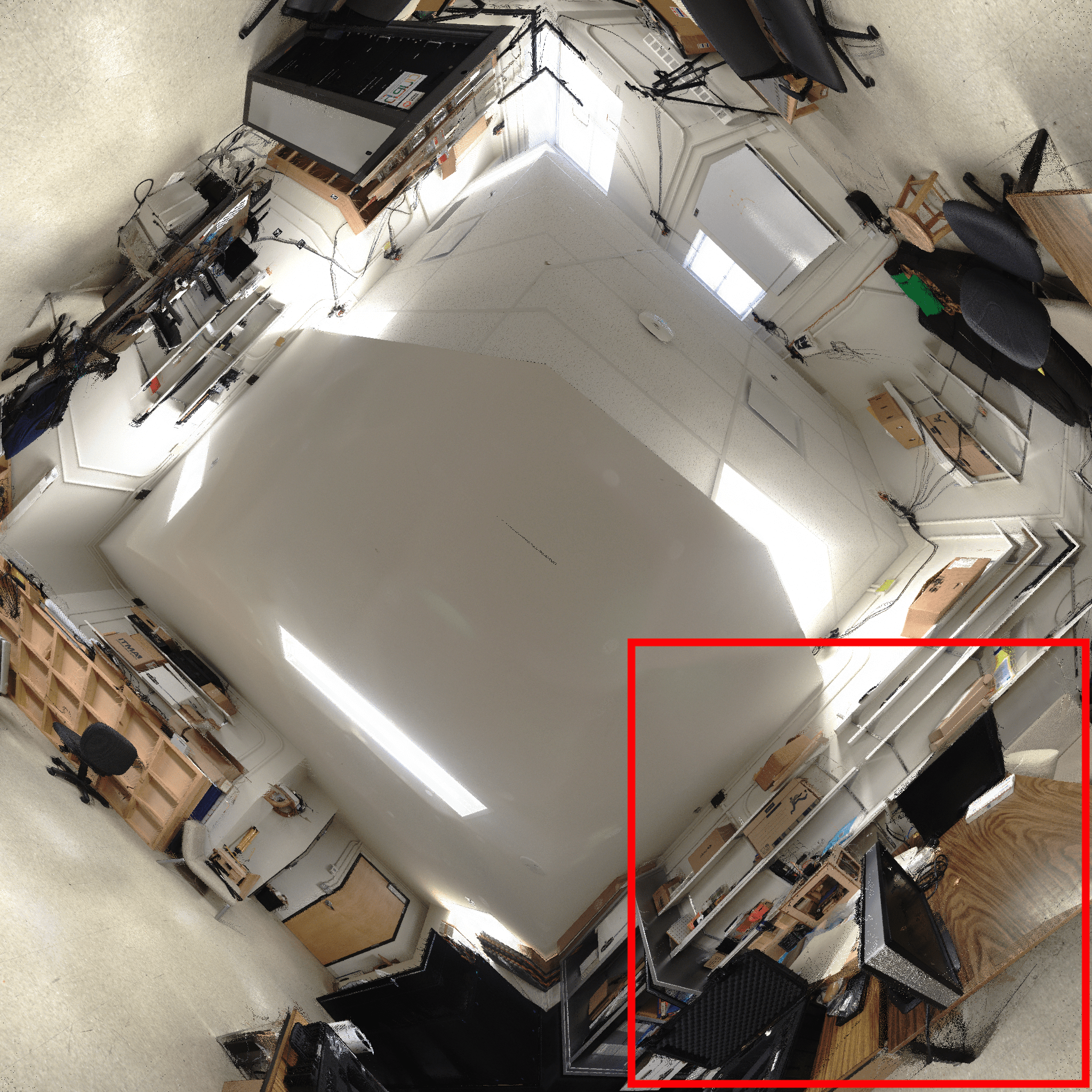}
         \caption{}
         \label{fig:5.2}
     \end{subfigure}
     \begin{subfigure}[b]{0.29\textwidth}
         \centering
         \includegraphics[width=\textwidth]{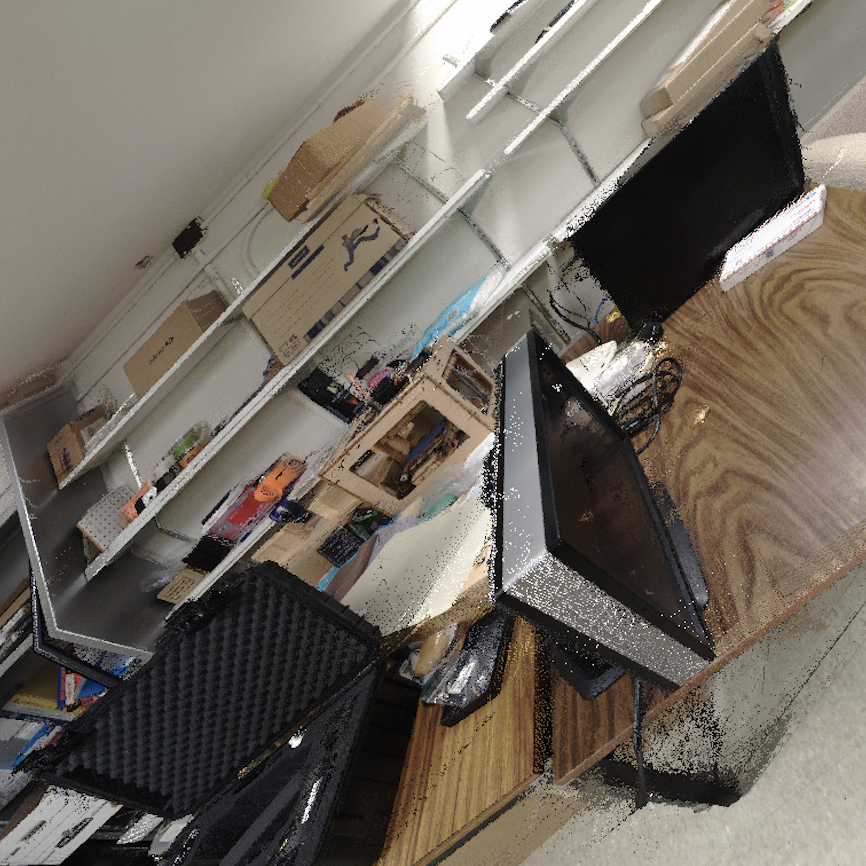}
         \caption{}
         \label{fig:5.3}
     \end{subfigure}
     \begin{subfigure}[b]{0.29\textwidth}
         \centering
         \includegraphics[width=\textwidth]{tablefloor.png}
         \caption{}
         \label{fig:a2}
     \end{subfigure}
        \caption[]{Irradiance map and its aliasing: (a) A 2048x2048 irradiance octahedral map of the lab; (b) the enlarged highlighted area of (a); the area is poorly captured by the scanner because the scanner is placed much above the table and far from the area; (c) the reconstruction output has lots of aliasing near the floor position.}
        \label{fig:5.4}
\end{figure}


\bigskip
\noindent \textbf{Hierarchical Representation}
Finer octahedral maps have 2048x2048 resolution, and the coarser maps have 128x128 resolution. The octahedral maps are generated offline. 
The unstructured real-world data is highly noisy even after data processing, so it is not surprising that the irradiance maps have aliasing. For example, the area below the table and near the floor (Figure~\ref{fig:5.2}). The aliasing is due to the fact that the scanner is placed above the table and hardly sees the area near the floor. Moreover, the aliasing of probe data is exaggerated on the rendering output (Figure~\ref{fig:5.4}).

\begin{figure}
     \centering
     \begin{subfigure}[b]{0.3\textwidth}
         \centering
         \includegraphics[width=\textwidth]{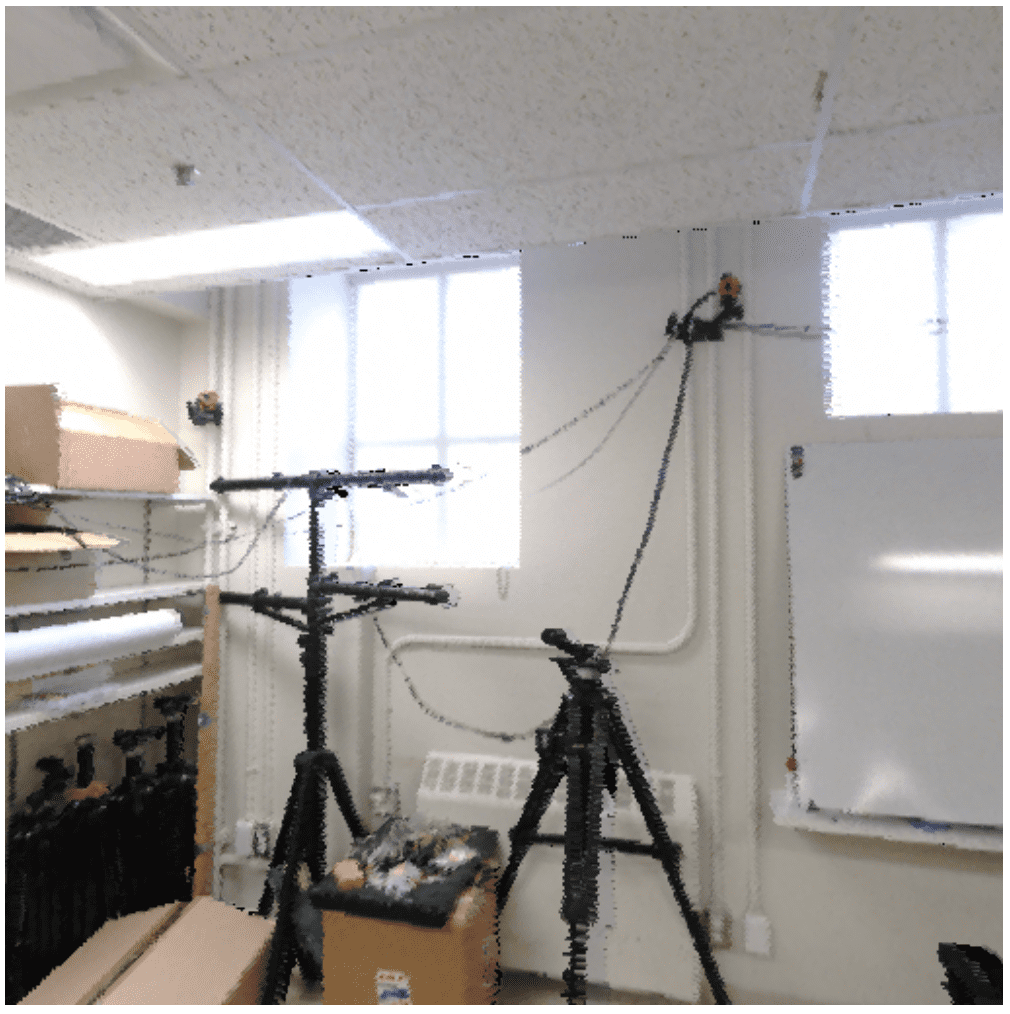}
         \caption{}
         \label{fig:58a}
     \end{subfigure}
     \begin{subfigure}[b]{0.3\textwidth}
         \centering
         \includegraphics[width=\textwidth]{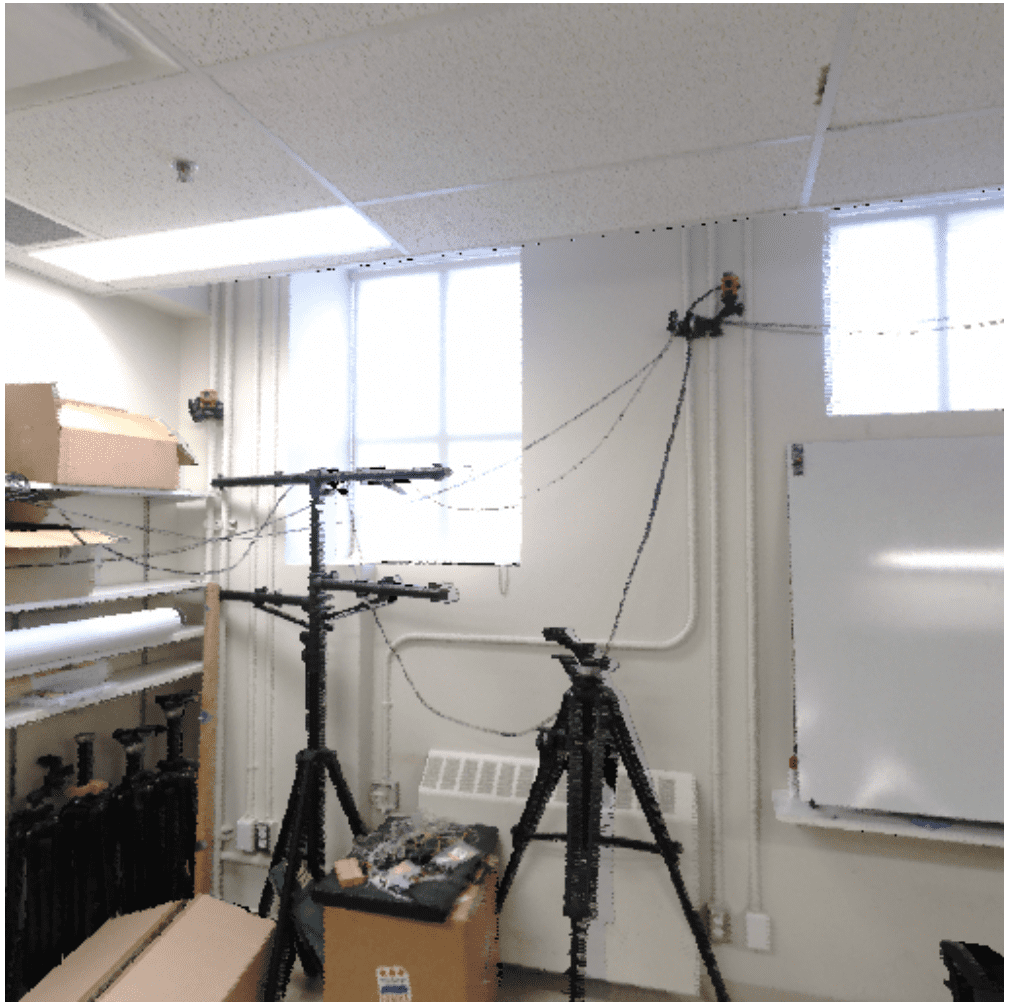}
         \caption{}
         \label{fig:58b}
     \end{subfigure}
     \begin{subfigure}[b]{0.3\textwidth}
         \centering
         \includegraphics[width=\textwidth]{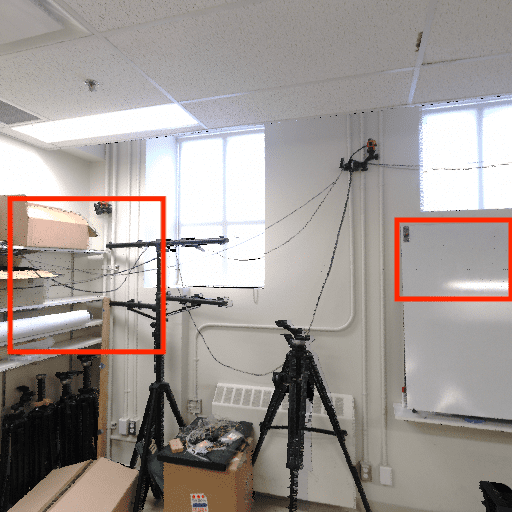}
         \caption{}
         \label{fig:58c}
     \end{subfigure}
        \caption[Scene reconstructions using 1024x1024, 2048x2048 probe data]{Scene reconstructions with different high-resolution probe data; the reconstruction
using 1024x1024 probe data has more jig jags, for example along the edge of the window; the reconstruction using 3072x3072 probe data has more aliasing (see highlighted area of figure c)the reconstructions use probe data with resolution: (a) 1024x1024 and 64x64; (b) 2048x2048 and 64x64; (c) 3072x3072 and 64x64.}
        \label{fig:5.8}
\end{figure}

\begin{figure}
     \centering
     \begin{subfigure}[b]{0.3\textwidth}
         \centering
         \includegraphics[width=\textwidth]{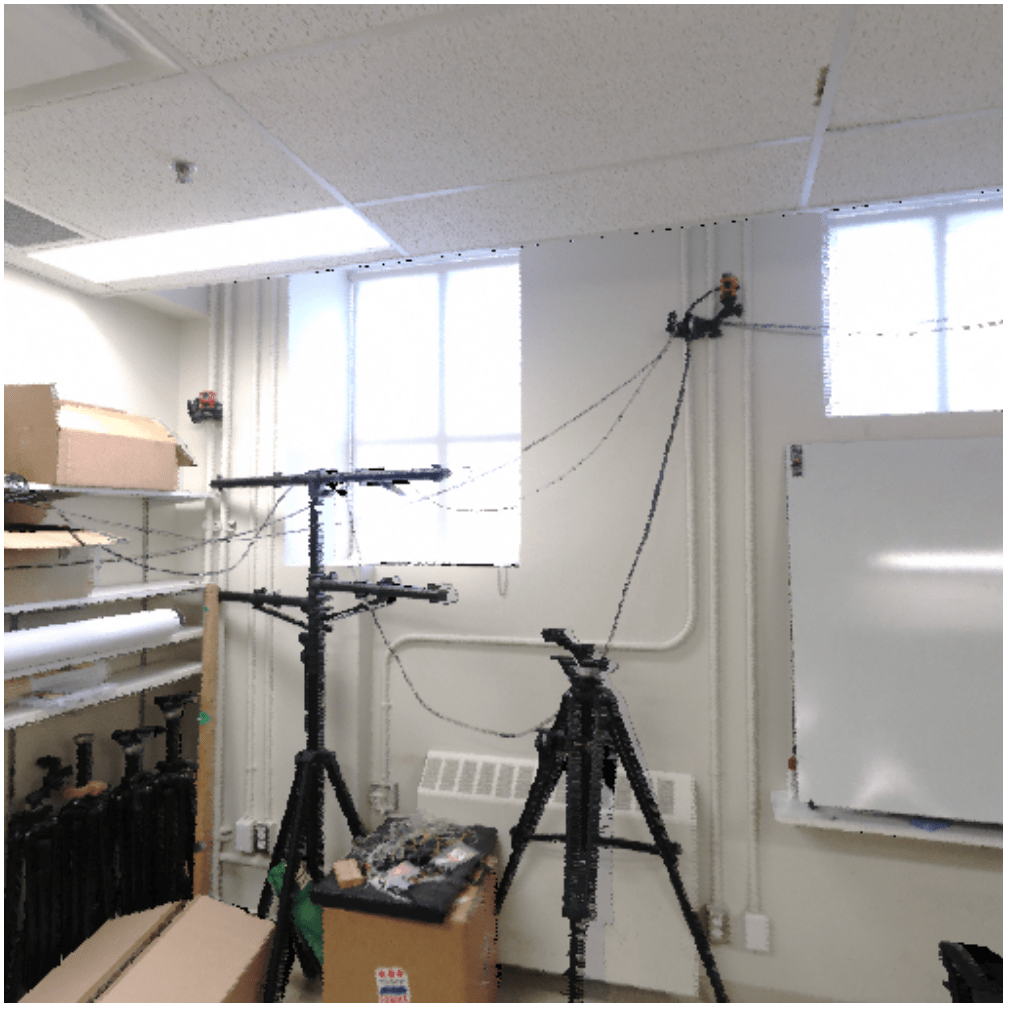}
         \caption{}
         \label{fig:a1}
     \end{subfigure}
     \begin{subfigure}[b]{0.3\textwidth}
         \centering
         \includegraphics[width=\textwidth]{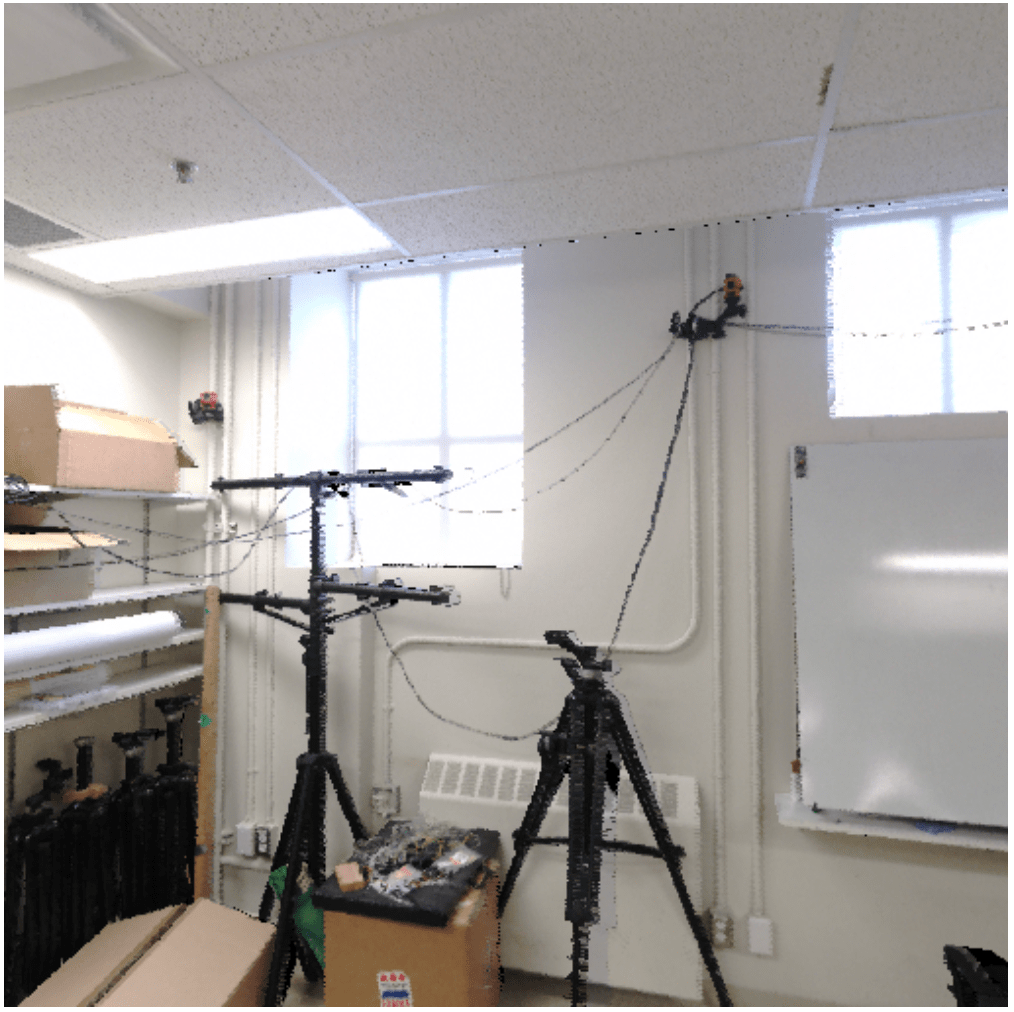}
         \caption{}
         \label{fig:a2}
     \end{subfigure}
        \caption[Scene reconstruction quality not affected by low-resolution probe data]{Low-resolution probe data do not affect the scene reconstruction quality; the reconstructions use probe data with resolution: (A) 2048x2048 and 64x64; (B) 2048x2048 and 128x128.}
        \label{fig:5.9}
\end{figure}

Increasing the resolution of octahedral maps will theoretically enhance the reconstruction quality. To verify the assumption, we generate probe data of 1024x1024, 2048x2048, and 3072x3072 for finer probe data and 64x64 and 128x128 for coarser probe data. Compared to reconstruction using 2048x2048 and 64x64 probe data, reconstruction using 1024x1024 and 64x64 probe data has more jig jags, especially on the edges of scene objects (Figure~\ref{fig:5.8}). 
Similarly, reconstruction using 3072x3072 and 64x64 probe data also leads to fewer jig jags. However, it has more aliasing than 2048x2048 and 64x64 probe data (Figure~\ref{fig:58a}, \ref{fig:58b}). Moreover, the experiments verify that low-resolution octahedral maps do not affect the reconstruction quality (Figure~\ref{fig:5.9}). It is the highest-resolution probe data that affects the rendering results. Using probe data with resolution 3072x3072 leads to more aliasing compared to 2048x2048. This is different from our previous assumption, but makes sense because more point cloud data are needed to fill the 3072x3072 octahedral maps. Adding more point cloud data can generate higher quality 3072x3072 octahedral maps and thus avoid the aliasing. But more data increases memory usage and computation cost.

\subsection{Reconstruction}
\begin{figure}[th]
\centering
\includegraphics[width=15cm]{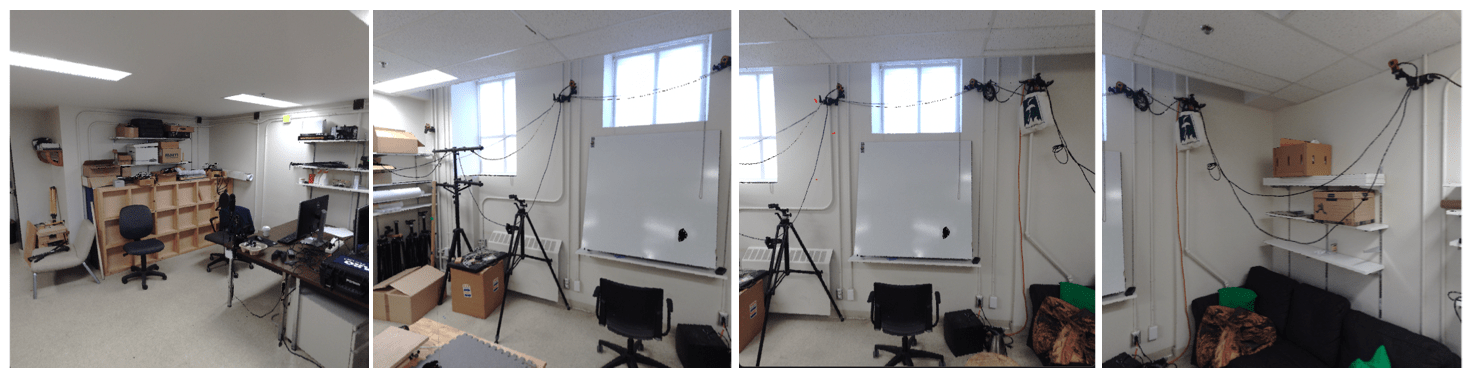}
\caption[Rendering]{Reconstruction of the scene using only 1 probe data; the eye is aligned with the probe's position.}
\label{fig:5.77}
\end{figure}

\textbf{Memory Efficient Representation}
Figure~\ref{fig:5.77} shows the reconstruction results when the eye is at the probe's position. The scene is reconstructed using a single probe. The results show that our probe-based technique is able to recover fine details and output high-quality synthesis. Using 2-level hierarchical representation with 2048x2048 and 128x128 resolutions, each probe involves five 2D images (Figure~\ref{fig:4.1}) and takes 24.375MB. 
Therefore, representing scene using probe data is highly memory efficient.

\begin{figure}[th]
\centering
\includegraphics[width=15cm]{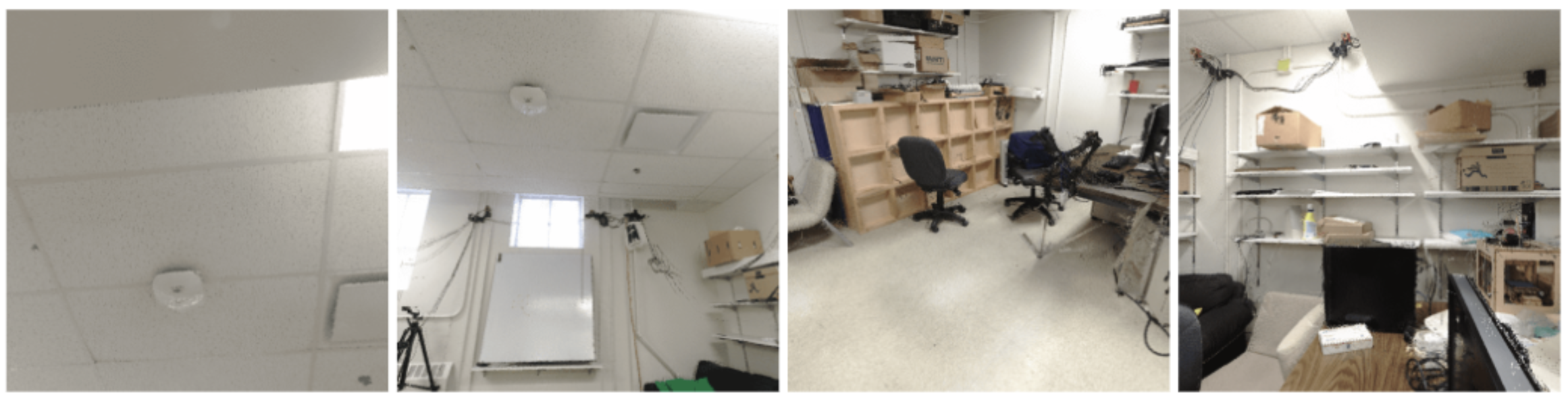}
\caption[Rendering time and scene complexity]{Rendering time does not grow linearly as the scene complexity increases; to render a frame, each takes: (a) 11.36ms; (b) 11.5ms; (c) 11.23ms; (d) 11.11ms.}
\label{fig:5.78}
\end{figure}

\bigskip
\noindent \textbf{Rendering Time Independent of Scene Complexity}
One significant advantage of using the probe representation is that rendering time is independent of scene complexity. Compared to the traditional rasterization method, our technique is more efficient.
This is because the probe data encodes all environment information into 2D images, which are rendered offline. Rasterization uses a z-buffer to find the closet point to shade and thus the computation cost grows linearly as the complexity and the size of the scene grows
Our experiment results verify this property. We set up experiments to render different parts of a room with data captured by different probes. As shown in Figure~\ref{fig:5.78}, the rendering time for different complexity of the scene is highly consistent, all around 11-12ms.

\subsection{Comparisons}
To evaluate our technique, we compare against Gaussian Splatting [\cite{kerbl_3d_2023}] for view synthesis. Both methods reconstruct the same lab scene but the input data are captured at different time of the day so the illumination looks different. The training material for Gaussian Splatting is a video taken by a phone's camera. When comparing the reconstruction results, we call both the scanner and the phone's camera as camera.

\begin{figure}[th]
\centering
\includegraphics[width=15cm]{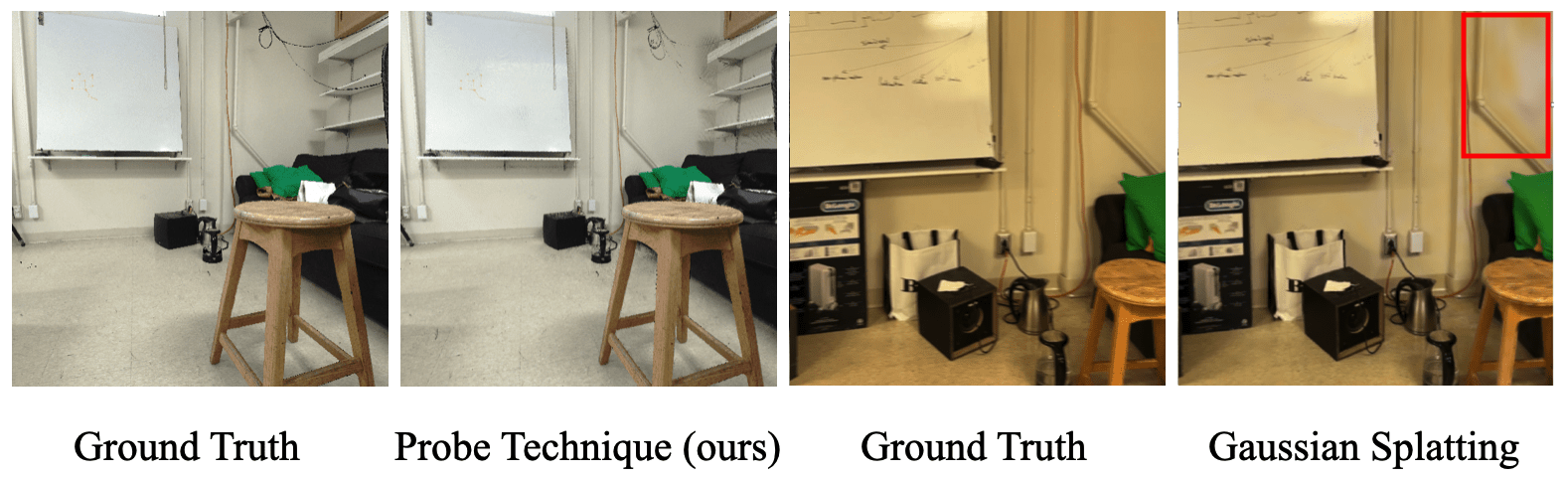}
\caption[compare]{Comparisons on test scene when the eye is aligned with the camera. Our method is able to recover fine details and even the reflection on the brink of the wooden chair. Gaussian Splatting produces blurry and inconsistent color around the wall area, as highlighted by the red box.}
\label{fig:5.5}
\end{figure}

\bigskip
\noindent \textbf{Eye Aligned with Camera}
When the eye aligns with a camera, 
the synthesized output looks nearly the same as what is captured by the camera (Figure~\ref{fig:5.5}). These near-identical outputs prove that our methodology can synthesize high-quality outputs. However, the reconstruction using Gaussian Splatting exhibits artifacts at the wall (see the artifacts highlighted by the red box Figure~\ref{fig:5.5}).  

\begin{figure}[H]
     \centering
     \begin{subfigure}[b]{1\textwidth}
         \centering
         \includegraphics[width=\textwidth]{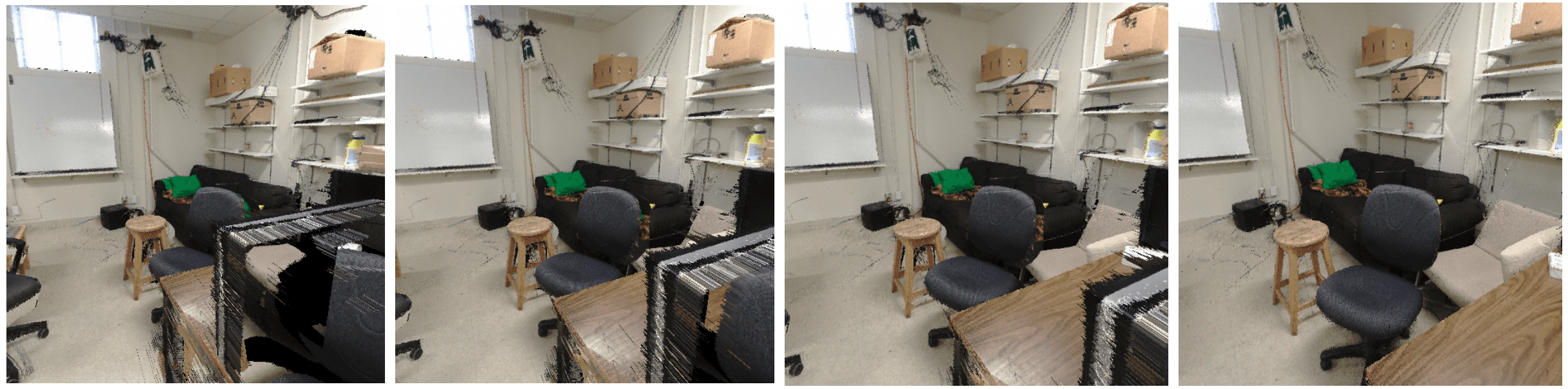}
         \caption{}
         \label{fig:5.19a}
     \end{subfigure}
     \begin{subfigure}[b]{1\textwidth}
         \centering
         \includegraphics[width=\textwidth]{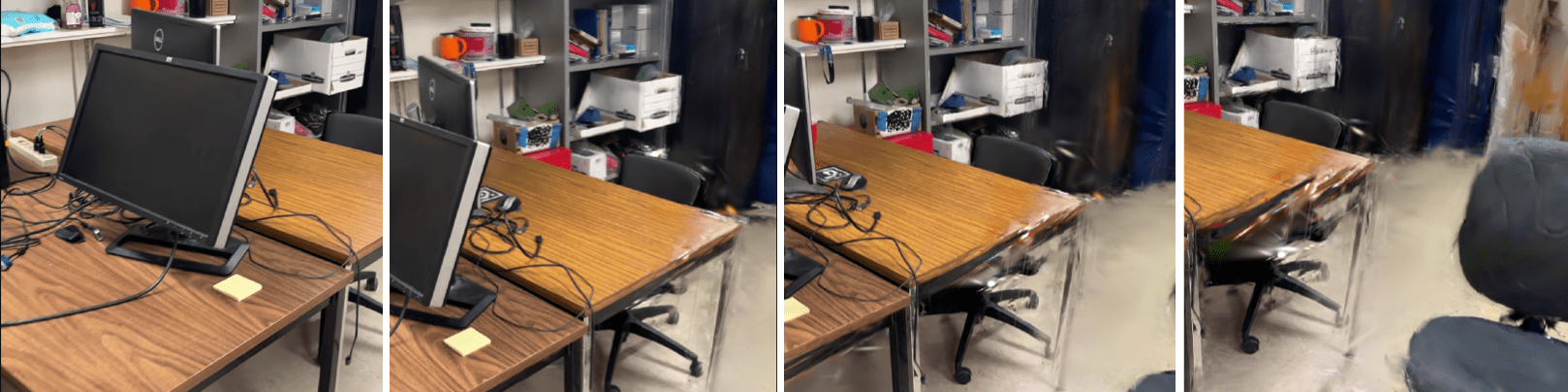}
         \caption{}
         \label{fig:5.19b}
     \end{subfigure}
        \caption[]{The reconstruction quality becomes worse as the eye moves away from the camera: (a) reconstruction using probe technique without simulated probe, the eye moves closer to the camera; (b) reconstruction using Gaussian splatting, the eye moves away from the camera.}
        \label{fig:move}
\end{figure}

\begin{figure}[H]
     \centering
     \begin{subfigure}[b]{0.35\textwidth}
         \centering
         \includegraphics[width=\textwidth]{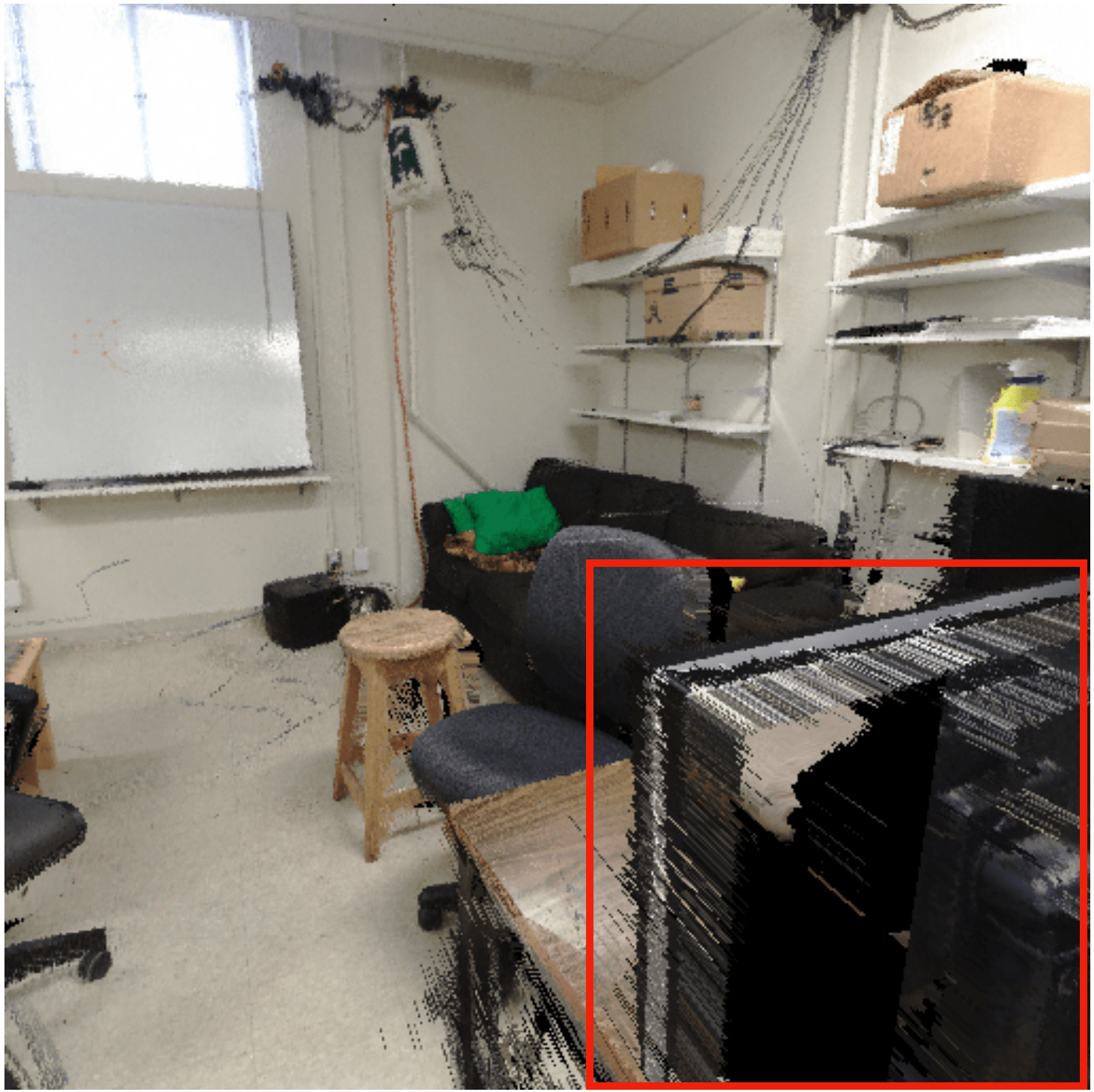}
         \caption{}
         \label{fig:5.19a}
     \end{subfigure}
     \begin{subfigure}[b]{0.35\textwidth}
         \centering
         \includegraphics[width=\textwidth]{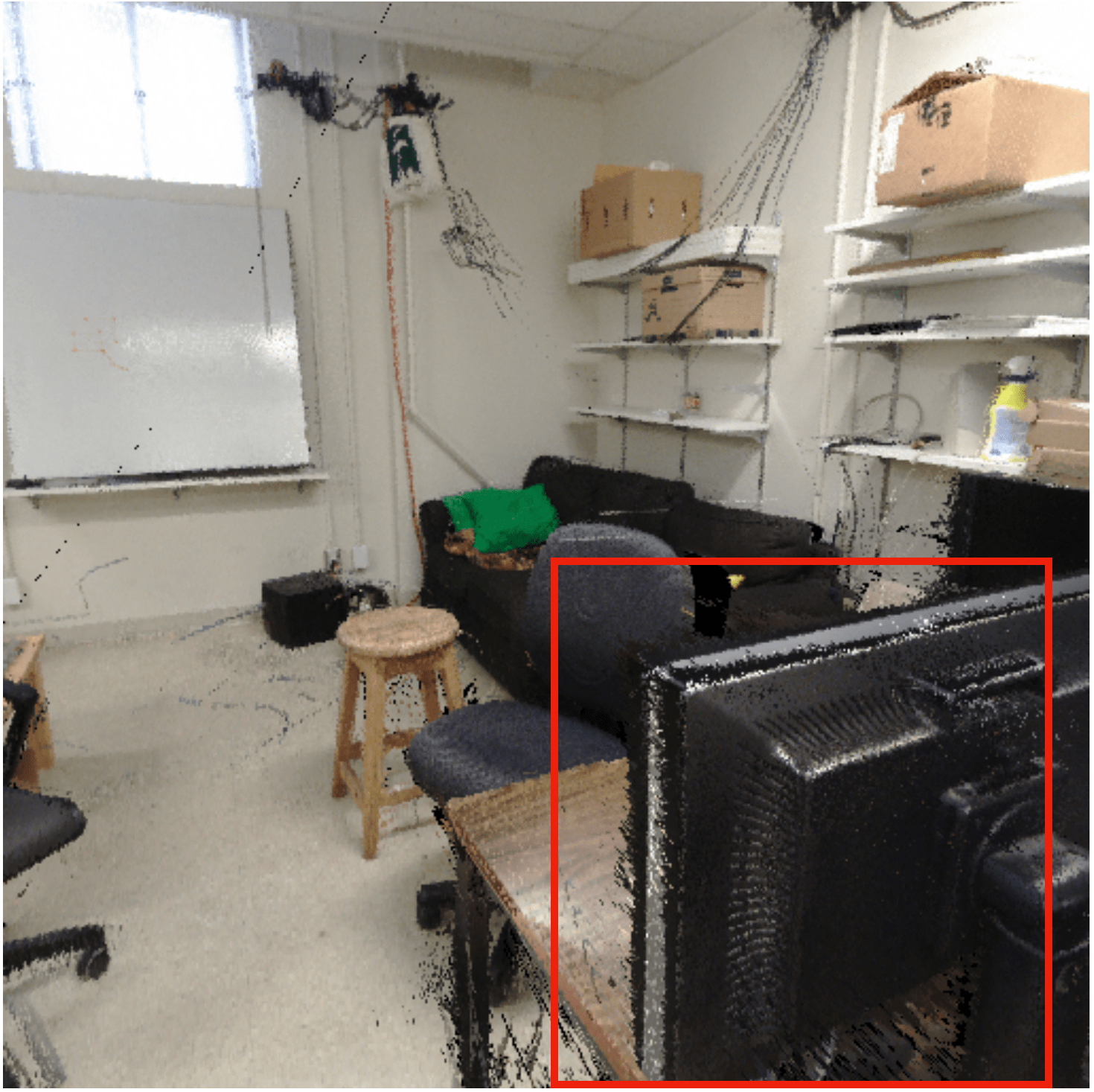}
         \caption{}
         \label{fig:5.19b}
     \end{subfigure}
        \caption[]{Reconstruction with simulated probe data can resolve the problem of missing data. Probe data are generated offline, so simulated data enables high-quality novel view synthesis in real-time: (a) reconstruction without simulated probe data; the highlighted area shows a missing part of a computer; (b) reconstruction with simulated probe data; the back of the computer is well rendered.}
        \label{fig:5.19}
\end{figure}

\bigskip
\noindent \textbf{Eye Not Aligned with Camera}
If the eye is not aligned with the probe, both techniques introduces artifacts.In particular, the reconstruction quality is worsen as the eye moves further away from the camera (Figure~\ref{fig:move}). However, our method can resolve this problem by using simulated probe data which is preprocessed offline (Figure~\ref{fig:5.19}), so we can achieve high quality real-time rendering. But to fix the problem in Gaussian Splatting, additional training is required and thus we can not render in real-time.

\section{Limitations and Future Work}

When multiple probes are used during reconstruction, the registered project point cloud data suffers from noise and outliers. One primary reason is that the same objects are captured from different angles, thus generating lots of redundant data. Moreover, real-world data is highly noisy and comes with many artifacts, such as outliers and low signal-to-noise. In addition, more probes introduce more noise and redundancy issues.
Our data processing does not denoise the point cloud data nor resolve data redundancy. So denoising the project point cloud either using machine learning or traditional denoising algorithms is interesting future work.

One direction is to apply machine learning to process the unstructured point cloud data. Machine learning can potentially be used to improve the quality of the probe data or transmit the data by encoding and decoding. We believe using machine learning for probe data transmission helps to reconstruct large-scale scenes.

Moreover, future work can include numerical analysis, which is important in computer graphics. Section {\ref{section7results}} gives a thorough visualization and analysis, but error values are not provided due to the fact that analyzing errors in real-world scenes is difficult. However, error analysis might be done by leveraging FARO software. Providing error analysis gives a more rigorous analysis of the artifacts produced by the probe-based technique. Therefore, it is recommended that the error analysis be added to future work.


\section{Conclusion}
In the paper, we have explored efficiently synthesizing a 3D scene using unstructured point cloud data captured from the real world instead of explicit geometry. 
We apply our framework to reconstruct a highly complex lab room. 
The results show that our probe-based reconstruction framework achieves to efficiently render a highly complex room-scale scene in real-time. One significant advantage of using probe data is that the rendering time is independent of the scene complexity. So we can efficiently reconstruct a scene of any complexity as long as high-quality probe data are available. Moreover, when rendering large-scale scenes, compressing
and streaming probe data is more efficient than using explicit scene geometry. 
When the eye moves away from the probes' positions, the reconstruction produces artifacts. We creatively resolve this problem by introducing simulated probe data. The simulated probe data enables high-quality rendering in real-time without introducing excessive data used for probe generation.
Our neural representation approach can potentially be applied to VR and AR applications.


\printbibliography

@String{Computing = "Computing" }

@String{Computer = "{IEEE} Computer" }

@inproceedings{gi-probes,
	address = {New York, NY, USA},
	series = {{I3D} '17},
	title = {Real-time global illumination using precomputed light field probes},
	isbn = {978-1-4503-4886-7},
	url = {https://doi.org/10.1145/3023368.3023378},
	doi = {10.1145/3023368.3023378},
	urldate = {2023-08-03},
	booktitle = {Proceedings of the 21st {ACM} {SIGGRAPH} {Symposium} on {Interactive} {3D} {Graphics} and {Games}},
	publisher = {Association for Computing Machinery},
	author = {McGuire, Morgan and Mara, Mike and Nowrouzezahrai, Derek and Luebke, David},
	month = feb,
	year = {2017},
	pages = {1--11},
}

@misc{fastnerf,
	title = {{FastNeRF}: {High}-{Fidelity} {Neural} {Rendering} at {200FPS}},
	shorttitle = {{FastNeRF}},
	url = {http://arxiv.org/abs/2103.10380},
	doi = {10.48550/arXiv.2103.10380},
	urldate = {2023-08-03},
	publisher = {arXiv},
	author = {Garbin, Stephan J. and Kowalski, Marek and Johnson, Matthew and Shotton, Jamie and Valentin, Julien},
	month = apr,
	year = {2021},
	note = {Number: arXiv:2103.10380
arXiv:2103.10380 [cs]},
}

@misc{neuralode,
  doi = {10.48550/ARXIV.2101.10994},
  
  url = {https://arxiv.org/abs/2101.10994},
  
  author = {Takikawa, Towaki and Litalien, Joey and Yin, Kangxue and Kreis, Karsten and Loop, Charles and Nowrouzezahrai, Derek and Jacobson, Alec and McGuire, Morgan and Fidler, Sanja},
  
  title = {Neural Geometric Level of Detail: Real-time Rendering with Implicit 3D Shapes},
  
  publisher = {arXiv},
  
  year = {2021},
  
  copyright = {Creative Commons Attribution Non Commercial Share Alike 4.0 International}
}

@misc{nerf,
	title = {{NeRF}: {Representing} {Scenes} as {Neural} {Radiance} {Fields} for {View} {Synthesis}},
	shorttitle = {{NeRF}},
	url = {http://arxiv.org/abs/2003.08934},
	doi = {10.48550/arXiv.2003.08934},
	publisher = {arXiv},
	author = {Mildenhall, Ben and Srinivasan, Pratul P. and Tancik, Matthew and Barron, Jonathan T. and Ramamoorthi, Ravi and Ng, Ren},
	month = aug,
	year = {2020},
	note = {Number: arXiv:2003.08934
arXiv:2003.08934 [cs]},
}

@misc{BungeeNeRF,
	title = {{BungeeNeRF}: {Progressive} {Neural} {Radiance} {Field} for {Extreme} {Multi}-scale {Scene} {Rendering}},
	shorttitle = {{BungeeNeRF}},
	url = {http://arxiv.org/abs/2112.05504},
	doi = {10.48550/arXiv.2112.05504},
	urldate = {2023-08-03},
	publisher = {arXiv},
	author = {Xiangli, Yuanbo and Xu, Linning and Pan, Xingang and Zhao, Nanxuan and Rao, Anyi and Theobalt, Christian and Dai, Bo and Lin, Dahua},
	month = may,
	year = {2023},
	note = {Number: arXiv:2112.05504
arXiv:2112.05504 [cs]},
}

@misc{wildnerf,
	title = {{NeRF} in the {Wild}: {Neural} {Radiance} {Fields} for {Unconstrained} {Photo} {Collections}},
	shorttitle = {{NeRF} in the {Wild}},
	url = {http://arxiv.org/abs/2008.02268},
	doi = {10.48550/arXiv.2008.02268},
	urldate = {2023-08-03},
	publisher = {arXiv},
	author = {Martin-Brualla, Ricardo and Radwan, Noha and Sajjadi, Mehdi S. M. and Barron, Jonathan T. and Dosovitskiy, Alexey and Duckworth, Daniel},
	month = jan,
	year = {2021},
	note = {Number: arXiv:2008.02268
arXiv:2008.02268 [cs]},
}

@misc{PlenOctrees,
	title = {{PlenOctrees} for {Real}-time {Rendering} of {Neural} {Radiance} {Fields}},
	url = {http://arxiv.org/abs/2103.14024},
	doi = {10.48550/arXiv.2103.14024},
	urldate = {2023-08-03},
	publisher = {arXiv},
	author = {Yu, Alex and Li, Ruilong and Tancik, Matthew and Li, Hao and Ng, Ren and Kanazawa, Angjoo},
	month = aug,
	year = {2021},
	note = {Number: arXiv:2103.14024
arXiv:2103.14024 [cs]},
}

@article{Blinn1TextureAR,
	title = {Texture and reflection in computer generated images},
	volume = {19},
	issn = {0001-0782},
	url = {https://dl.acm.org/doi/10.1145/360349.360353},
	doi = {10.1145/360349.360353},
	number = {10},
	urldate = {2023-08-07},
	journal = {Communications of the ACM},
	author = {Blinn, James F. and Newell, Martin E.},
	month = oct,
	year = {1976},
	pages = {542--547},
}

@article{musgrave,
  title={Light Probe Selection Algorithms for Real-Time Rendering of Light Fields},
  author={F Kenton Musgrave},
  year={1988},
  Journal = {Research Report No. RR-639, Dept. of Computer Science, Yale Univ}
}

@misc{aliev_neural_2020,
	title = {Neural {Point}-{Based} {Graphics}},
	url = {http://arxiv.org/abs/1906.08240},
	urldate = {2023-08-03},
	publisher = {arXiv},
	author = {Aliev, Kara-Ali and Sevastopolsky, Artem and Kolos, Maria and Ulyanov, Dmitry and Lempitsky, Victor},
	month = apr,
	year = {2020},
	note = {Number: arXiv:1906.08240
arXiv:1906.08240 [cs]},
}

@inproceedings{carr_reconstruction_2001,
	address = {New York, NY, USA},
	series = {{SIGGRAPH} '01},
	title = {Reconstruction and representation of {3D} objects with radial basis functions},
	isbn = {978-1-58113-374-5},
	url = {https://doi.org/10.1145/383259.383266},
	doi = {10.1145/383259.383266},
	urldate = {2023-08-03},
	booktitle = {Proceedings of the 28th annual conference on {Computer} graphics and interactive techniques},
	publisher = {Association for Computing Machinery},
	author = {Carr, J. C. and Beatson, R. K. and Cherrie, J. B. and Mitchell, T. J. and Fright, W. R. and McCallum, B. C. and Evans, T. R.},
	month = aug,
	year = {2001},
	pages = {67--76},
}

@misc{burov_dynamic_2021,
	title = {Dynamic {Surface} {Function} {Networks} for {Clothed} {Human} {Bodies}},
	url = {http://arxiv.org/abs/2104.03978},
	urldate = {2023-08-03},
	publisher = {arXiv},
	author = {Burov, Andrei and Nießner, Matthias and Thies, Justus},
	month = aug,
	year = {2021},
	note = {Number: arXiv:2104.03978
arXiv:2104.03978 [cs]},
}

@misc{genova_local_2020,
	title = {Local {Deep} {Implicit} {Functions} for {3D} {Shape}},
	url = {http://arxiv.org/abs/1912.06126},
	doi = {10.48550/arXiv.1912.06126},
	urldate = {2023-08-07},
	publisher = {arXiv},
	author = {Genova, Kyle and Cole, Forrester and Sud, Avneesh and Sarna, Aaron and Funkhouser, Thomas},
	month = jun,
	year = {2020},
	note = {arXiv:1912.06126 [cs]},
}

@misc{sitzmann_deepvoxels_2019,
	title = {{DeepVoxels}: {Learning} {Persistent} {3D} {Feature} {Embeddings}},
	shorttitle = {{DeepVoxels}},
	url = {http://arxiv.org/abs/1812.01024},
	doi = {10.48550/arXiv.1812.01024},
	urldate = {2023-08-03},
	publisher = {arXiv},
	author = {Sitzmann, Vincent and Thies, Justus and Heide, Felix and Nießner, Matthias and Wetzstein, Gordon and Zollhöfer, Michael},
	month = apr,
	year = {2019},
	note = {Number: arXiv:1812.01024
arXiv:1812.01024 [cs]},
}

@article{majercik_scaling_2021,
	title = {Scaling {Probe}-{Based} {Real}-{Time} {Dynamic} {Global} {Illumination} for {Production}},
	volume = {10},
	language = {en},
	number = {2},
	author = {Majercik, Zander and Marrs, Adam and Spjut, Josef and McGuire, Morgan},
	year = {2021},
}

@article{majercik_dynamic_2019,
	title = {Dynamic {Diffuse} {Global} {Illumination} with {Ray}-{Traced} {Irradiance} {Fields}},
	volume = {8},
	language = {en},
	number = {2},
	author = {Majercik, Zander and Guertin, Jean-Philippe and Nowrouzezahrai, Derek and McGuire, Morgan},
	year = {2019},
}

@inproceedings{agarwal_building_2009,
	title = {Building {Rome} in a day},
	url = {https://ieeexplore.ieee.org/abstract/document/5459148},
	doi = {10.1109/ICCV.2009.5459148},
	urldate = {2023-10-19},
	booktitle = {2009 {IEEE} 12th {International} {Conference} on {Computer} {Vision}},
	author = {Agarwal, Sameer and Snavely, Noah and Simon, Ian and Seitz, Steven M. and Szeliski, Richard},
	month = sep,
	year = {2009},
	note = {ISSN: 2380-7504},
	pages = {72--79},
}

@misc{rematas_urban_2021,
	title = {Urban {Radiance} {Fields}},
	url = {http://arxiv.org/abs/2111.14643},
	doi = {10.48550/arXiv.2111.14643},
	urldate = {2023-10-19},
	publisher = {arXiv},
	author = {Rematas, Konstantinos and Liu, Andrew and Srinivasan, Pratul P. and Barron, Jonathan T. and Tagliasacchi, Andrea and Funkhouser, Thomas and Ferrari, Vittorio},
	month = nov,
	year = {2021},
	note = {arXiv:2111.14643 [cs]},
}

@misc{tancik_block-nerf_2022,
	title = {Block-{NeRF}: {Scalable} {Large} {Scene} {Neural} {View} {Synthesis}},
	shorttitle = {Block-{NeRF}},
	url = {http://arxiv.org/abs/2202.05263},
	doi = {10.48550/arXiv.2202.05263},
	urldate = {2023-10-19},
	publisher = {arXiv},
	author = {Tancik, Matthew and Casser, Vincent and Yan, Xinchen and Pradhan, Sabeek and Mildenhall, Ben and Srinivasan, Pratul P. and Barron, Jonathan T. and Kretzschmar, Henrik},
	month = feb,
	year = {2022},
	note = {arXiv:2202.05263 [cs]},
}

@misc{turki_mega-nerf_2022,
	title = {Mega-{NeRF}: {Scalable} {Construction} of {Large}-{Scale} {NeRFs} for {Virtual} {Fly}-{Throughs}},
	shorttitle = {Mega-{NeRF}},
	url = {http://arxiv.org/abs/2112.10703},
	doi = {10.48550/arXiv.2112.10703},
	urldate = {2023-10-19},
	publisher = {arXiv},
	author = {Turki, Haithem and Ramanan, Deva and Satyanarayanan, Mahadev},
	month = mar,
	year = {2022},
	note = {arXiv:2112.10703 [cs]},
}

@misc{rebain_derf_2020,
	title = {{DeRF}: {Decomposed} {Radiance} {Fields}},
	shorttitle = {{DeRF}},
	url = {http://arxiv.org/abs/2011.12490},
	doi = {10.48550/arXiv.2011.12490},
	urldate = {2023-10-19},
	publisher = {arXiv},
	author = {Rebain, Daniel and Jiang, Wei and Yazdani, Soroosh and Li, Ke and Yi, Kwang Moo and Tagliasacchi, Andrea},
	month = nov,
	year = {2020},
	note = {arXiv:2011.12490 [cs]},
}

@article{neff_donerf_2021,
	title = {{DONeRF}: {Towards} {Real}-{Time} {Rendering} of {Compact} {Neural} {Radiance} {Fields} using {Depth} {Oracle} {Networks}},
	volume = {40},
	issn = {0167-7055, 1467-8659},
	shorttitle = {{DONeRF}},
	url = {http://arxiv.org/abs/2103.03231},
	doi = {10.1111/cgf.14340},
	number = {4},
	urldate = {2023-10-19},
	journal = {Computer Graphics Forum},
	author = {Neff, Thomas and Stadlbauer, Pascal and Parger, Mathias and Kurz, Andreas and Mueller, Joerg H. and Chaitanya, Chakravarty R. Alla and Kaplanyan, Anton and Steinberger, Markus},
	month = jul,
	year = {2021},
	note = {arXiv:2103.03231 [cs]},
	pages = {45--59},
}

@inproceedings{reiser_kilonerf_2021,
	address = {Montreal, QC, Canada},
	title = {{KiloNeRF}: {Speeding} up {Neural} {Radiance} {Fields} with {Thousands} of {Tiny} {MLPs}},
	isbn = {978-1-66542-812-5},
	shorttitle = {{KiloNeRF}},
	url = {https://ieeexplore.ieee.org/document/9710464/},
	doi = {10.1109/ICCV48922.2021.01407},
	language = {en},
	urldate = {2023-10-19},
	booktitle = {2021 {IEEE}/{CVF} {International} {Conference} on {Computer} {Vision} ({ICCV})},
	publisher = {IEEE},
	author = {Reiser, Christian and Peng, Songyou and Liao, Yiyi and Geiger, Andreas},
	month = oct,
	year = {2021},
	pages = {14315--14325},
}

@misc{hedman_baking_2021,
	title = {Baking {Neural} {Radiance} {Fields} for {Real}-{Time} {View} {Synthesis}},
	url = {http://arxiv.org/abs/2103.14645},
	doi = {10.48550/arXiv.2103.14645},
	urldate = {2023-10-19},
	publisher = {arXiv},
	author = {Hedman, Peter and Srinivasan, Pratul P. and Mildenhall, Ben and Barron, Jonathan T. and Debevec, Paul},
	month = mar,
	year = {2021},
	note = {arXiv:2103.14645 [cs]},
}

@misc{garbin_fastnerf_2021,
	title = {{FastNeRF}: {High}-{Fidelity} {Neural} {Rendering} at {200FPS}},
	shorttitle = {{FastNeRF}},
	url = {http://arxiv.org/abs/2103.10380},
	doi = {10.48550/arXiv.2103.10380},
	urldate = {2023-10-19},
	publisher = {arXiv},
	author = {Garbin, Stephan J. and Kowalski, Marek and Johnson, Matthew and Shotton, Jamie and Valentin, Julien},
	month = apr,
	year = {2021},
	note = {arXiv:2103.10380 [cs]},
}

@article{kerbl_3d_2023,
	title = {{3D} {Gaussian} {Splatting} for {Real}-{Time} {Radiance} {Field} {Rendering}},
	volume = {42},
	issn = {0730-0301, 1557-7368},
	url = {https://dl.acm.org/doi/10.1145/3592433},
	doi = {10.1145/3592433},
	language = {en},
	number = {4},
	urldate = {2024-01-04},
	journal = {ACM Transactions on Graphics},
	author = {Kerbl, Bernhard and Kopanas, Georgios and Leimkuehler, Thomas and Drettakis, George},
	month = aug,
	year = {2023},
	pages = {1--14},
}

\appendix

\section{Algorithms} 
\begin{algorithm}
	\caption{RayMarching(($L_p, r_p$), $(X, \omega)$)} 
	\begin{algorithmic}[1]
            \State $R \leftarrow$ projectRayOntoProbeData$(X, \omega)$
		\For {$P$ on projected ray $R$} 
			\State $distanceFromProbeToP \leftarrow$ texelFetch$(L_p, r_p, P)$
			\State $directionFromProbeToPIn3D$ $\leftarrow$ octahedralToWorldSpace$(P)$
			\State $distanceToIntersection$ $\leftarrow$ distanceFromProbeToPIn3D$((X, \omega), directionFromProbeToPIn3D)$

               \If{$distanceFromProbeToP < distanceToIntersection$}
                  \State return (true, $P$);
                \EndIf
		\EndFor
	\end{algorithmic} 
\label{fig:alg1}
\end{algorithm}

\begin{algorithm}
	\caption{traceOneProbe(($L_p, r_p$), $(X, \omega), probeOrigin$)} 
	\begin{algorithmic}[5]

           \If{$X == probeOrigin$}
              \State uv = projectOntoProbeData($\omega$));
              \State irradiance = texelFetch($L_p, r_p$, uv).color
            \Else
                \State irradiance = traceRaySegments($L_p, r_p$, probeOrigin)
            \EndIf
	\end{algorithmic} 
\label{fig:alg5}
\end{algorithm}

\begin{algorithm}
	\caption{traceOneRaySegment(($L_p, r_p$), $(X, \omega)$, $t_0, t_1$)} 
	\begin{algorithmic}[2]
            \State $rayStart \leftarrow$ findPositionInProbeSpace$((X, \omega), t_0)$
            \State $rayEnd \leftarrow$ findPositionInProbeSpace$((X, \omega), t_1)$
            \State texCoord $\leftarrow$ findTexCoord(rayStart)
		\While {true} 
                \State lowResResult, $uv, uv_{end}$ = lowResolutionTracing(($L_p^{lowRes}, 
                r_p^{lowRes}$), texCoord, $(X, \omega)$)
               \If{lowResResult is false}
                  \State return (MISS, None);
                \Else
                    \State (result, uv) $\leftarrow$ highResolutionTracing(($L_p^{highRes}, r_p^{highRes}$), $uv, uv_{end}$, $(X, \omega)$)
                    \If{result != MISS}
                        \State return (result, uv)
                    
                    \EndIf
                    \State texCoord $\leftarrow$ adjustTexCoord(texCoord)
                \EndIf
		\EndWhile
            \State return (MISS, None)
	\end{algorithmic} 
\label{fig:alg2}
\end{algorithm}

\begin{algorithm}
	\caption{lowResolutionTracing(($L_p, r_p$), texCoord, $(X, \omega)$)} 
	\begin{algorithmic}[3]
            \State $P \leftarrow$ findUVCoord(texCoord)
            \State $R \leftarrow$ projectRayOntoProbeData$(X, \omega)$
		\For {$P$ on projected ray $R$} 
			\State $distanceFromProbeToP \leftarrow$ texelFetch$(L_p, r_p, P).distance$
			\State $directionFromProbeToPIn3D$ $\leftarrow$ octahedralToWorldSpace$(P)$
			\State $distanceToIntersection$ $\leftarrow$ distanceFromProbeToPIn3D$((X, \omega), directionFromProbeToPIn3D)$

               \If{$distanceFromProbeToP < distanceToIntersection$}
                    \State $P_{end} \leftarrow findEndOfPixel(P)$ 
                  \State return (true, $P, P_{end}$);
                \EndIf
		\EndFor
        \State return (false, $P$, None);
	\end{algorithmic} 
\label{fig:alg3}
\end{algorithm}

\begin{algorithm}[H]
	\caption{highResolutionTracing(($L_p, r_p$), uv, $uv_{end}, (X, \omega)$)} 
	\begin{algorithmic}[4]
            \State $P \leftarrow$ uv
		\For {$P < uv_{end}$} 
			\State $distanceFromProbeToP \leftarrow$ texelFetch$(L_p, r_p, P).distance$
			\State $directionFromProbeToPIn3D$ $\leftarrow$ octahedralToWorldSpace$(P)$
			\State $distanceToIntersection$ $\leftarrow$ distanceFromProbeToPIn3D$((X, \omega), directionFromProbeToPIn3D)$

               \If{$distanceFromProbeToP < distanceToIntersection$}
                    \State normal $\leftarrow texelFetch(L_p, r_p, P).normal$
                    \If{normal $\cdot \omega$ < 0 }
                        \State return (HIT, P);
                    \Else
                        \State return (UNKNOWN, P);
                    \EndIf
                \EndIf
		\EndFor
        \State return (MISS, P);
	\end{algorithmic} 
\label{fig:alg4}
\end{algorithm}

\section{Lab room}
\begin{figure}[th]
\centering
\includegraphics[width=13cm]{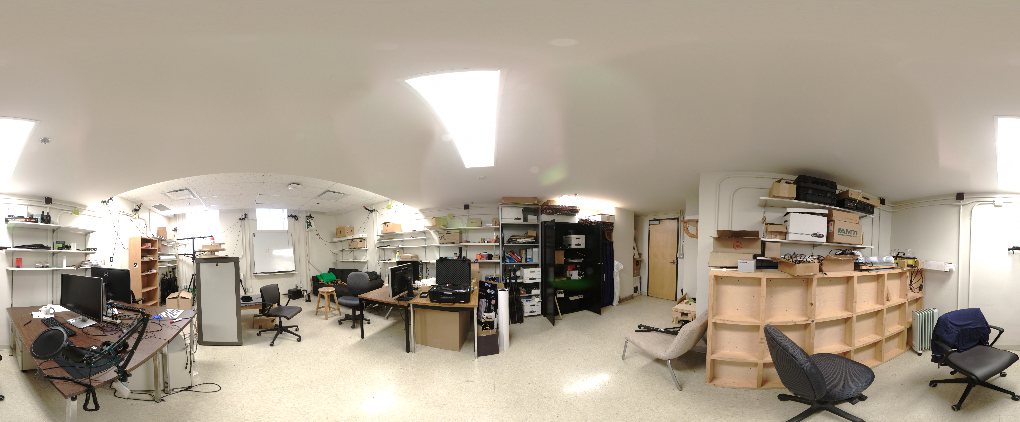}
\caption[Panoramic picture of the lab]{A panoramic picture of the lab.}
\label{fig:5.0}
\end{figure}

\begin{figure}[th]
\centering
\includegraphics[width=12cm]{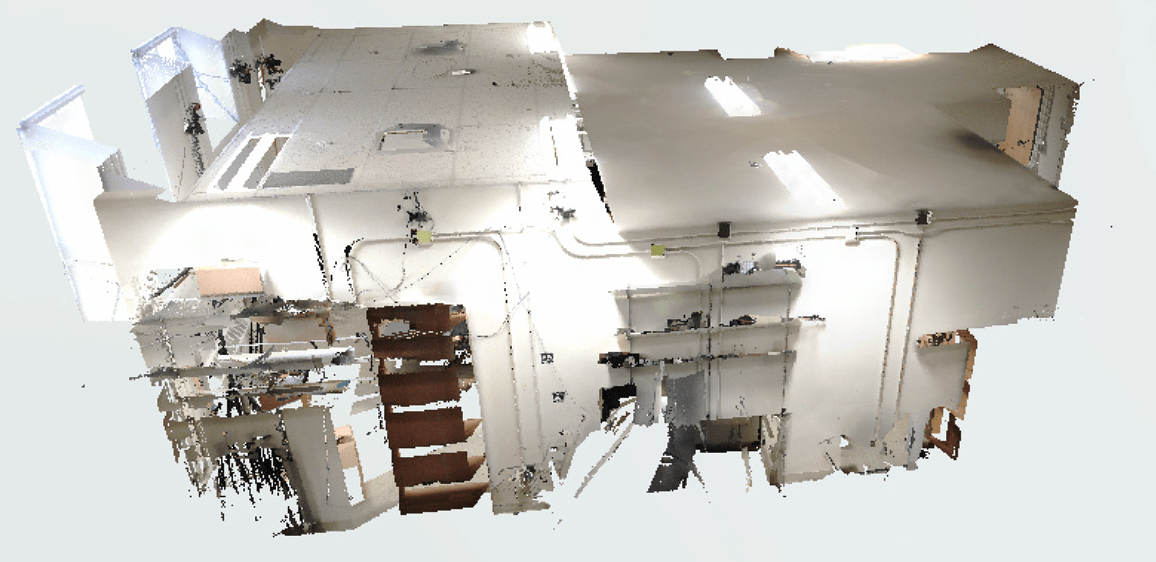}
\caption[Project point cloud data]{We create a project point cloud data that consists of all scans. The figure visualizes the project point cloud after colorization, registration and color balancing.}
\label{fig:3.8}
\end{figure}


\end{document}